\documentclass[12pt, draftclsnofoot, onecolumn]{IEEEtran}

\usepackage{caption}
\usepackage{clrscode}
\usepackage{bbm}
\usepackage{amsfonts}
\usepackage[dvips]{graphicx}
\usepackage{times}
\usepackage{cite}
\usepackage{amsmath}
\usepackage{array}
\usepackage{amssymb}

\usepackage{stfloats}
\usepackage{slashbox}
\usepackage{graphicx}
\usepackage{subfigure}
\usepackage{footnote}
\usepackage{amsthm}
\usepackage{booktabs}
\usepackage{array}
\usepackage{algorithm}
\usepackage{algorithmic}
\usepackage{subeqnarray}
\usepackage{cases}
\usepackage{threeparttable}
\usepackage{color}
\usepackage{multirow}
\usepackage{xcolor}
\usepackage{epstopdf}
\usepackage{bm}
\usepackage{dsfont}
\usepackage{enumerate}
\usepackage{float}
\usepackage{hyperref}

\normalsize

\begin{document}

\title{On the Fundamental Limits of MIMO Massive Access Communication}
\author{Fan Wei~\IEEEmembership{Student Member,~IEEE}, Yongpeng Wu,~\IEEEmembership{Senior Member,~IEEE}, \\
Wen Chen,~\IEEEmembership{Senior Member,~IEEE}, Yanlin Geng,~\IEEEmembership{Member,~IEEE}, 
       \\ and~Giuseppe Caire,~\IEEEmembership{Fellow,~IEEE}

\thanks{This paper was presented in part at IEEE ICC 2019. Y. Wu and W. Chen are the corresponding authors of this paper.}

\thanks{F. Wei and W. Chen are with Shanghai Institute of Advanced Communications and Data Sciences, the Department of Electronic
Engineering, Shanghai Jiao Tong University, Minhang 200240, China (E-mail: weifan89@sjtu.edu.cn; wenchen@sjtu.edu.cn).}

\thanks{Y. Wu is with the Department of Electronic Engineering,
Shanghai Jiao Tong University, Minhang 200240, China (e-mail: yongpeng.wu@sjtu.edu.cn)}

\thanks{Y. Geng is with State Key Lab of ISN, Xidian University, 
Xi’an, China (e-mail: ylgeng@xidian.edu.cn)}

\thanks{G. Caire is with Institute for Telecommunication Systems, Technical University Berlin, Einsteinufer 25,
10587 Berlin, Germany (Email: caire@tu-berlin.de).}

}

\maketitle

\begin{abstract}
The multiple access channel (MAC) with many-user is a general model for massive machine type communications. In this paradigm, the number of users may be comparable or even larger than the coding blocklength $n$. In contrast, classical MAC often assumes fixed and small number of the users. In this paper, we consider the massive access channel with multiple antennas system, where the base station (BS) with multiple receiving antennas serves the users in a single cell. The magnitude of users is assumed to grow unbounded with $n$. We investigate the achievable region of MIMO massive access channel, where among the total users, an unknown subset referred to active users may transmit data periodically. The asymptotic active user identification cost is also quantified. With the theoretical analysis, it was shown that given finite number of the receiving antennas, the individual rate for each user can be formulated as the sum rate multiplied by some specific factors, which correspond to the allocation of sum capacity. The successive decoding does not apply due to the interferences from growing unbounded users. Theoretical analysis shows that successive decoding works only when the number of receiving antennas goes to infinity with the increasing codelength.
\end{abstract}


%
\IEEEpeerreviewmaketitle

\section{Introduction}
\IEEEPARstart{M}{assive} machine type communication (mMTC) has found its applications to various practical scenarios, such as wireless sensor networks, Device to Device (D2D) communication, and Internet of
Things (IoT). A typical communication scenario would be a large pool of devices, with the magnitude of which may range from thousands to millions are simultaneously served by base station (BS) within single or multiple cells. Meanwhile, the users may send signals with a certain probability due to the sporadic traffic. When no data need to be delivered, the devices are kept in a sleep mode in order to save energy. Such communication scenarios have arisen two key differences from conventional multiple access (MAC) channels. In the first place, the conventional MAC assumes a fixed and small number of users such that the coding blocklength goes to infinity before the number of users operates in a same way. In contrast, in massive access communications, the number of active users may be comparable or even larger than the coding blocklength. In the second place, since the devices transmit data sporadically with certain probabilities, the decoder must be able to identify the set of active users before decoding, whereas in conventional MAC accurate user activity information is presumed.

To accommodate the such features, a many-access channel (MnAC) model was introduced in~\cite{Chen17,Chen13,Chen14}. The key difference between MnAC and conventional MAC is that in MnAC, the number of active users may be comparable or even larger than the coding blocklength. The setup suggests a different approach to characterize the fundamental limits of MnAC. Mathematically, for a given number of users $k_{n}$, which is a function of coding blocklength $n$, $\lim_{k\rightarrow\infty}\lim_{n\rightarrow\infty}f(k,n)\neq \lim_{n\rightarrow\infty}f(k_{n},n)$ for some functions $f(\cdot)$. The definition of MnAC model was introduced in~\cite{Chen13}. There, it was shown random coding with Feinstein’s threshold decoding suffices to achieve the symmetric capacity of Gaussian MnAC, when $k_{n}$ grows sublinearly with $n$. The linearly growing case was further investigated in~\cite{Chen17}, where the user identification cost for random access was also characterized. The technique for proof was based on Gallager's error exponent analysis~\cite{Gallager68}.

The models in MnAC~\cite{Chen17,Chen13,Chen14} presume single antenna at both transceivers. On the other hand, wireless networks with multiple antennas~\cite{Telatar99,Goldsmith03,Shin03,Marzetta10} simultaneously serving massive users is a promising $5$G technology. In this paper, we investigate the achievable region of massive access channels with multiple antennas. The setup is akin to that in many-access channels. The difference is that multiple antenna arrays can be deployed at both transceivers. We characterize the asymmetric capacity of multiple-input multiple-output (MIMO) massive access channels, as well as the asymptotic user identification cost with random devices access. For the active devices identification, since active devices often constitute only a small subset of the potential ones, the sparse user recovery can be formulated as a compressed sensing problem~\cite{Donoho06,Jin11,Aksoylar17,Scarlett17}. Thus, the user identification cost would be a by product of the asymptotic bound for sparse recovery in MIMO massive access channels. We derive the asymptotic user identification bound through concentration inequalities~\cite{Boucheron13} which are related to the information density for signature symbols. From the viewpoint of communications, the derived compressed sensing bounds are further interpreted.

The asymmetric capacity for MIMO massive access channels when the set of active devices are known perfectly by the receiver is also investigated. We show that the rates can be formulated as the sum capacity of the system multiplied by some codebook size related factors that corresponds to the rate allocations. The asymptotic behavior of asymmetric capacity when the number of users goes to infinity is also characterized. Combining the results for user identification and data transmission, we finally obtain the finite dimension region of MIMO massive random access channels. In contrast to conventional MAC, where the achievable rates are determined by signal to interference plus noise (SINR), and are achieved by successive decoding at receiver~\cite{Cover06}, the successive interference cancellation may not be applicable in massive access channels given finite number of the receiving antennas. The underlying reason is by the growing unbounded user interferences in massive access channels~\cite{Chen17}. As a last contribution of this paper, we further investigate the conditions for the number of receiving antennas required to be deployed when successive decoding works in MIMO massive access channels.

In addition to MnAC model, some recent works consider the user number $K$ may grow unbounded with the number of available resources $N$, such that the overloading factor $\beta=K/N$ converges to some constant~\cite{Shental17,Zaidel18,Le18,Mai-Le18}. The spectral efficiency for non-orthogonal multiple access (NOMA) with large system limit is investigated through the analysis of limiting spectral density of the spreading matrix~\cite{Shental17}. Closed-form expression are further derived for optimal and linear minimum-mean-square error receivers, respectively~\cite{Zaidel18}. The interesting results has suggested the superiority of regular sparse NOMA compared with the irregular one and also the dense randomly-spread code-division multiple-access, where the optimal decoding becomes prohibitive. Large system limit with fading scenario is further considered in~\cite{Le18}.

It should be noted that the user random access is absent in the formulation of above models. For massive MIMO system, the achievable rates for massive connectivity with random user access are characterized in~\cite{Liu18II}. In contrast to this paper, the works in~\cite{Liu18II} assumes the number of receiving antennas may vary with user numbers such that the ratio between two quantities is kept fixed. Some practical schemes for user activity detection are further investigated in~\cite{Liu18I,Chen18}. In~\cite{Polyanskiy17}, random access code is defined such that the achievability bounds can be compared with existing schemes such as ALOHA, Code Division Multiple Access (CDMA), treating interference as noise (TIN), and etc. Some coding schemes~\cite{Ordentlich,Amalladinne18,Fengler19} are considered later in order to achieve the above random code bound. A recent work~\cite{Kowshik19} also considers the combining model of MnAC with the code definitions in~\cite{Polyanskiy17}, the bounds on the optimal required energy-per-bit is derived but with no considering on random access. For the MIMO system, a new scaling law has been investigated showing that given sufficient number of receiving antennas, the number of stably estimated active users may exceed the conventional one which is constrained by compressed sensing based user detection~\cite{Haghighatshoar19}.

Unless otherwise noted, we use the following notational conventions: the lowercase letters $x$, bold lowercase letters $\mathbf{x}$, and bold uppercase letters $\mathbf{X}$ are used to denote scalars, column vectors, and matrices, respectively. We use $(\cdot)^{\dagger}$ to denote complex conjugate and transpose for matrix and $\mbox{Tr}\{\cdot\}$ to denote the trace operator of matrix. The notion $\binom{n}{k}$ denotes the binomial coefficient of $n$ choose $k$. The binary entropy function is denoted by $H_{2}(p)=-p\log p-(1-p)\log(1-p)$. {We use $\mathcal{CN}(x;\tau, \upsilon)$ to denote $x$ follows complex Gaussian distribution with the mean $\tau$ and variance $\upsilon$.} The asymptotic notations $O(\cdot)$, $o(\cdot)$, $\Theta(\cdot)$, $\Omega(\cdot)$, and $\omega(\cdot)$ follows a standard meaning~\cite{Jin11}. The $\log(\cdot)$ functions are taken with natural base throughout this paper.


\section{System Model}
\subsection{MIMO Massive Access Channels}
\begin{figure}
  \centering
  \includegraphics[width=3.5in]{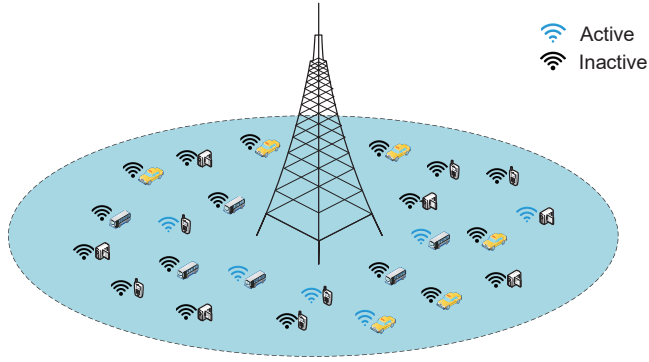}
  \caption{MIMO massive random access network}\label{Fig.1}
\end{figure}

We consider a MIMO massive access channels with $\ell_{n}$ potential user equipments (UE) shown as in Fig.~\ref{Fig.1}, where the receiver has $N_{R}$ received antennas while the UEs are equipped with $N_{T}$ transmit antennas. In the massive multiple access, the number of users is comparable or even larger than the coding blocklength $n$. Therefore, unless dealing with the user identification problem, we denote $\ell_{n}$ the number of potential users, and $k_{n}$ the number of active users on average, while the user transmission probability is given by $\alpha_{n}=k_{n}/\ell_{n}$. Since the UEs may be active or inactive depending on their data traffics, the uplink transmission can be divided into two phases. In the training phase, for the sake of user activity identification, the active UEs send signature signals to the base station (BS) with a signature length $n_{0}$. Thereafter, the active UEs send codewords with a coding blocklength $n-n_{0}$ in the data transmission phase.

Let $\mathcal{T}=\{1,2,\ldots,\ell_{n}\}$ denote the total user set, and $\mathcal{A}\subseteq \mathcal{T}$ denote the active UE set with the average size $|\mathcal{A}|$ equals $k_{n}=O(n)$, the received signature signals at BS can be written as
\begin{align}
    \label{01a}\mathbf{y}(i)&=\sum_{k \in \mathcal{A}}\mathbf{H}_{k}(i)\mathbf{s}_{k}(i)+\mathbf{z}(i)\\
    \label{01b}&=\sum_{k \in \mathcal{T}}\mathbf{H}_{k}(i)\mathbf{s}_{k}(i)x_{k}+\mathbf{z}(i)\\
    \label{01c}&=\mathbf{H}(i)\mathbf{S}(i)\mathbf{x}+\mathbf{z}(i), \,\, i = 1,2,\ldots,n_{0}
\end{align}
where $\mathbf{H}(i)=\big[\mathbf{H}_{1}(i),\mathbf{H}_{2}(i),\ldots,\mathbf{H}_{\ell_{n}}(i)\big]$, and $\mathbf{S}(i)=\mbox{diag} \big\{\mathbf{s}_{1}(i),\mathbf{s}_{2}(i),\ldots,\mathbf{s}_{\ell_{n}}(i)\big\}$. The matrix $\mathbf{H}_{k}(i) \in \mathbb{C}^{N_{R}\times N_{T}}$ denotes the channel gain from UE $k$, $\mathbf{s}_{k}(i) \in \mathbb{C}^{N_{T}}$ represents the signature symbols from UE $k$, which is supposed to be independent and identically distributed (i.i.d.) according to complex Gaussian distribution, and $\mathbf{z}(i) \in \mathbb{C}^{N_{R}}$ is the complex Gaussian noise with each entry from $\mathcal{CN}(0,1)$. The Bernoulli vector $\mathbf{x}=[x_{1},x_{1},\cdots,x_{\ell_{n}}]^{T}$ follows since $x_{k}$s are the set of i.i.d. Bernoulli variables with probability $\alpha_{n}=k_{n}/\ell_{n}$. Given $x_{k} = 1$, UE $k$ is active otherwise $x_{k}=0$.

For each channel use, the coefficient for each antenna within the channel matrix $\mathbf{H}_{k}(i)$ can be formulated as
\begin{equation}\label{01d}
  h_{l,m}^{(k)}(i) = \alpha_{l,m}^{(k)}(i)\beta_{k}^{\frac{1}{2}}, \,\, l \in [1,N_{R}], \, m \in [1,N_{T}],
\end{equation}
where the small scale fading factor $\alpha_{l,m}^{(k)}(i)$ follows standard normal distribution $\mathcal{CN}(0,1)$, and varies through different channel uses. On the other hand, the large scale fading $\beta_{k}$ remains constant during the $n$ channel uses.
For channel state information (CSI), we assume perfect knowledge of channel can be obtained at the receiver (CSIR) while only channel distribution information is available at the transmitter (CDIT).

Let $\mathbf{X}=\mathbf{x}\otimes \mathbf{I}_{n_{0}}$, which denotes the Kronecker product of Bernoulli vector $\mathbf{x}=\{x_{1},x_{1},\cdots x_{\ell_{n}}\}^{T}$ and the identity matrix, the received signal in~\eqref{01c} can be rewritten as
\begin{equation}\label{02}
  \mathbf{Y}^{n_{0}}=\mathbf{H}^{n_{0}}\mathbf{S}^{n_{0}}\mathbf{X}+\mathbf{Z}^{n_{0}}
   =\mathbf{\Phi}^{n_{0}} \mathbf{X}+\mathbf{Z}^{n_{0}},
\end{equation}
where $\mathbf{Y}^{n_{0}}=\big[\mathbf{y}(1),\mathbf{y}(2),\ldots,\mathbf{y}(n_{0})\big]_{N_{R}\times n_{0}}$, the channel and signature matrix are given by
\begin{equation}\label{03}
  \mathbf{H}^{n_{0}} =   \begin{bmatrix}
                   \mathbf{H}(1), & \mathbf{H}(2), & \cdots, & \mathbf{H}(n_{0}) \\
                 \end{bmatrix}_{N_{R}\times n_{0}N_{T}\ell_{n}},
\end{equation}
and
\begin{equation}\label{04}
  \mathbf{S}^{n_{0}} = \begin{bmatrix}
                 \mathbf{S}(1) &  &  &  \\
                  & \mathbf{S}(2) &  &  \\
                  & & \ddots &  \\
                  & & & \mathbf{S}(n_{0}) \\
               \end{bmatrix}_{n_{0}N_{T}\ell_{n}\times n_{0}\ell_{n}},
\end{equation}
respectively. Given $\ell_{n}\gg k_{n}$, the detection of sparse vector $\mathbf{x}$ can be formulated as a compressed sensing problem, where $\mathbf{\Phi}^{n_{0}}=\mathbf{H}^{n_{0}}\mathbf{S}^{n_{0}}$ is the sensing matrix.

The received data signals follows
\begin{equation}\label{05}
  \check{\mathbf{y}}(i) = \sum_{k \in \mathcal{T}}\mathbf{H}_{k}(i)\check{\mathbf{s}}_{k}(i)+\mathbf{z}(i),\,\,
  i = n_{0}+1,n_{0}+2,\ldots,n
\end{equation}
where $\check{\mathbf{s}}_{k}(i)\in \mathbb{C}^{N_{T}}$ denotes the transmitted codewords from user $k$, and $\check{\mathbf{s}}_{k}(i)=\mathbf{0}$ when user $k$ is inactive. The channel and Gaussian noise follow the analogous definitions as in Eq.~\eqref{01c}.

\subsection{Codes Construction and Message-Length Capacity}
In massive access channels, as the number of active users $k_{n}$ scales linearly with the codelength, the individual user rate would approach zero with fixed transmit powers and number of antennas due to the growing unbounded user interferences~\cite[Section 14.3]{Cover06}. As a consequence, the traditional notion of rate defined by $R =\log M/n \rightarrow0$ (where $M$ denotes the codebook size), i.e., bits per channel use becomes a less meaningful performance metric. To circumvent the above problem, we consider a codelength related notion for transmission rate, which is a referred to message-length capacity~\cite{Chen17}. The following definitions introduce the asymmetric codes
construction as well as the notion of message-length capacity for MIMO massive access channels.

\emph{Definition 1 (\textcolor[rgb]{0,0,1}{Codes Construction with Asymmetric Rate})}:  Let $\mathcal{S}_{k}$ and $\mathcal{Y}$ be the alphabet of input symbol for user $k$ and output symbol of the channel, respectively. An $(M_{1}, M_{2},\ldots, M_{\ell_{n}},n)$ code for MIMO massive access channels $(\mathcal{S}_{1} \times \mathcal{S}_{2} \times ... \times \mathcal{S}_{\ell_{n}}, P_{Y|S_{1},...,S_{\ell_{n}}}, \mathcal{Y})$ consists of

\begin{itemize}
  \item $\ell_{n}$ sets of integers $\mathcal{W}_{k} \in \{0,1,2,...,M_{k}\}$ called \emph{message sets}.
  \item The set of \emph{encoding functions} $\mathcal{E}_{k}: \mathcal{W}_{k}\rightarrow \mathcal{S}^{N_{T}\times(n-n_{0})}_{k} $ for every user $k$. For the set of active users, the transmitted codewords per channel use $\mathbf{s}_{k}\in \mathbb{C}^{N_{T}\times1}$ should satisfy the following power constraint:
\begin{equation*}
  p_{k}-\delta \leq \mathrm{Tr}\big(\mathbf{Q}_{k}\big) \leq p_{k},
\end{equation*}
where $\delta>0$ is an arbitrary small number and $p_{k} = \Theta(1)$, i.e., a constant independent of the codelength $n$, and $\mathbf{Q}_{k} = \mathbb{E}\big\{\mathbf{s}_{k}\mathbf{s}_{k}^{\dag}\big\} $ denotes the covariance matrix for codewords $\mathbf{s}_{k}$. For each user $k$, the symbol $\mathbf{s}_{k}$ are assumed to be generated i.i.d. according to the Gaussian distribution $\mathcal{CN}(\mathbf{0};\mathbf{Q}_{k})$.
  \item A \emph{decoding function} $\mathcal{D}: \mathcal{Y}^{N_{R}\times(n-n_{0})} \rightarrow \mathcal{W}_{1} \times \mathcal{W}_{2} \times ... \times \mathcal{W}_{\ell_{n}} $, which is a deterministic mapping that assigns a sequence of estimated messages $(\mathcal{W}_{1}, \mathcal{W}_{2}, ... ,\mathcal{W}_{\ell_{n}})$ to each received vector $\check{\mathbf{Y}}$.
\end{itemize}

The average error probability of the $(M_{1},M_{2},\ldots, M_{\ell_{n}},n)$ code is given by
\begin{equation}\label{06}
  P_{e}^{(n)} = \mathrm{Pr}\big\{\mathcal{D}(\check{\mathbf{Y}}^{(n-n_{0})}) \neq (\mathcal{W}_{1}, \mathcal{W}_{2}, \ldots ,\mathcal{W}_{\ell_{n}})\big\},
\end{equation}
where the messages $(\mathcal{W}_{1}, \mathcal{W}_{2}, \ldots ,\mathcal{W}_{\ell_{n}})$ are generated independently over the message set, i.e.,
\begin{equation}\label{07}
  \mathrm{Pr}\{\mathcal{W}_{k}=w\}=\left\{
                           \begin{array}{ll}
                             1-\alpha_{n}, & \hbox{$w=0$;} \\
                             \alpha_{n}/ M_{k}, & \hbox{$w\in \mathcal{W}_{k} \backslash \{0\}$.}
                           \end{array}
                         \right.
\end{equation}

\emph{Definition 2 (Asymptotically Achievable Message-Length)}: Given the set of functions $R_{k}(\cdot)$ that map natural numbers to some positive value, we say the message-length $R_{k}(n)$ are asymptotically achievable if there exists a sequence of $(\lceil\exp(R_{1}(n))\rceil,\lceil\exp(R_{2}(n))\rceil,\ldots, \lceil\exp(R_{\ell_{n}}(n))\rceil,n)$ codes in the sense of Definition $1$ such that the average error probability $P_{e}^{(n)}$ vanishes as $n \rightarrow \infty$.

The capacity region for MIMO massive access channels can be the closure sets of all achievable rate tuples $\big(R_{1}(n),R_{2}(n),\ldots,R_{k_{n}}(n)\big)$ for the $k_{n}$ active users. However, as $n$ becomes larger, the dimension of this rate tuple also increases since the number of users now grows unbounded with $n$. Thus, to avoid an increasing dimension of capacity region, we define the finite dimension region of MIMO massive access channel by noting that the message set size $M_{k}$ may not necessarily different from each other.

\emph{Definition 3 (Finite Dimension Message-Length Region)}: Let $K_{j}$ denote the number of users with message set size $\exp\{V_{j}(n)\}$, i.e., with message-length rate $R_{k}(n)=V_{j}(n)$ for $k\in \mathcal{A}$, where $V_{j}(n)$ are the set of ``distinct'' message-length rates in the system. {Then by grouping together the rates $R_{k}(n)$ that are equal to each other, the $J$-dimensional region of MIMO massive access channel is defined as the closure of the convex hull of all $\big(V_{1}(n),V_{2}(n),\ldots, V_{J}(n)\big)$, such that if $\sum_{j=1}^{J}K_{j}V_{j}(n)$ is upper bounded by the sum rate of the system, the error probability goes to zero asymptotically.}

For the finite and denumerable message set size, the dimension $J$ was expected to be independent of codelength $n$. Thus, in \emph{Definition 3}, a finite dimension region is formulated through the achievable rates for the given number of sustainable users.


\section{Main Results}
\begin{figure}
  \centering
  \includegraphics[width=3.5in]{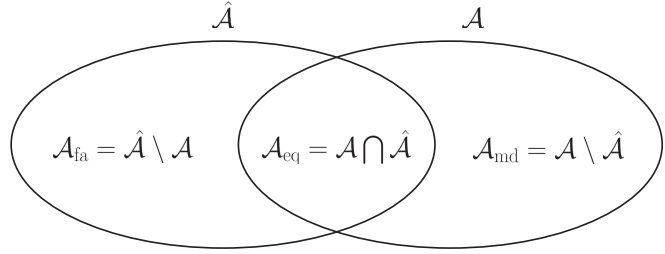}
  \caption{The relationship between sets $\mathcal{A}$ and $\hat{\mathcal{A}}$, where $|\mathcal{A}|=|\hat{\mathcal{A}}|=k_{n}$, $|\mathcal{A}_{\mathrm{md}}|=|\mathcal{A}_{\mathrm{fa}}|=i$.}\label{Fig.2}
\end{figure}

In this subsection, the main results of this paper are stated by the following theorems. For the UE identification in the training phase, the BS outputs a set $\hat{\mathcal{A}}$ with size $|\hat{\mathcal{A}}|=k_{n}$, which corresponds to the estimation of true set $\mathcal{A}$. Due to the random access nature, the size of active user set $\mathcal{A}$ may vary randomly in every transmission. Let $k_{\mathcal{A}}=|\mathcal{A}|$ be the random variable and $\delta>0$, by Chebyshev's inequality,
\begin{equation}\label{07a}
  \mathrm{Pr}\big\{|k_{\mathcal{A}}-k_{n}|\geq\delta n\big\} \leq \frac{V_{k_{\mathcal{A}}}}{\delta^{2} n^{2}},
\end{equation}
where $V_{k_{\mathcal{A}}}$ denotes the variance of random variable $k_{\mathcal{A}}$. Eq.~\eqref{07a} indicates that for finite $V_{k_{\mathcal{A}}}$, the probability that $k_{\mathcal{A}}$ deviates from the average number $k_{n}$ vanishes asymptotically as the codelength increases. Therefore in this paper, we assume that the decoded set $\hat{\mathcal{A}}$ has almost the same cardinality with $\mathcal{A}$ on average. Fig.~\ref{Fig.2} illustrates the set relationship between $\mathcal{A}$ and $\hat{\mathcal{A}}$, where $\mathcal{A}_{\mathrm{eq}}$ denotes the correct decoded set, $\mathcal{A}_{\mathrm{fa}}$ denotes the false alarm set, and $\mathcal{A}_{\mathrm{md}}$ corresponds to the misdetection set. Define the error decode probability as
\begin{equation}\label{08}
  P^{(\ell)}_{e}=\mathrm{Pr}\{\hat{\mathcal{A}}\neq \mathcal{A}\}.
\end{equation}
To find the limits for UE identification, we assume $\ell\rightarrow\infty$ and let the other parameters change with $\ell$. \emph{Theorem 1} provides the asymptotic UE identification cost for the centralized detection schemes.

\emph{Theorem 1 (UE Identification Cost for Massive Random Access Channels)}: Denote the total number of users as $\ell$ and the number of active users as $k_{\ell}$. Suppose $\ell$ and $k_{\ell}$ satisfy the following condition
\begin{equation}\label{09}
  \lim_{\ell\rightarrow\infty}\ell e^{-\varrho k_{\ell}} = 0
\end{equation}
for all $\varrho>0$, which indicates that $k_{\ell}\rightarrow \infty$ as $\ell\rightarrow\infty$. Let $i=|\mathcal{A}_{\mathrm{md}}|$, if the asymptotic number of UE identification cost for centralized detection is given by
\begin{equation}\label{11}
  n_{0}\geq (1+\epsilon)\cdot \max_{i=1,\ldots,k_{\ell}}  \frac{\log\binom{\ell-k_{\ell}}{i}}{
  \mathbb{E}_{\mathbf{H}}\Big\{\log\det\big(\mathbf{I}_{N_{R}}+\sum\limits_{k \in \mathcal{A}_{\mathrm{md}}}
  \mathbf{H}_{k}\mathbf{Q}_{k}\mathbf{H}^{\dagger}_{k}\big)\Big\}}
\end{equation}
for some $\epsilon>0$, the error probability can be made $P^{(\ell)}_{e}\rightarrow 0$ as $\ell$ goes to infinity.

Conversely, when the asymptotic number of UE identification cost
\begin{equation}\label{12}
  n_{0}\leq (1-\epsilon)\cdot \max_{i=1,\ldots,k_{\ell}} \frac{\log\binom{\ell-k_{\ell}+i}{i}}{
  \mathbb{E}_{\mathbf{H}}\Big\{\log\det\big(\mathbf{I}_{N_{R}}+\sum\limits_{k \in \mathcal{A}_{\mathrm{md}}}
  \mathbf{H}_{k}\mathbf{Q}_{k}\mathbf{H}^{\dagger}_{k}\big)\Big\}},
\end{equation}
for some $\epsilon>0$, we have $P^{(\ell)}_{e}\rightarrow 1$ as $\ell$ goes to infinity.
\hfill $\square$

After training phase, the receiver has a perfect knowledge of user activity within the network. The next phase is to decode the data from active UEs. \emph{Theorem 2} states the asymmetric rate for MIMO massive multiple access channels when the receiver knowing perfectly the true active UE set $\mathcal{A}$.

\emph{Theorem 2 (Asymmetric Message-Length Rate for Massive Access Channels)}: For MIMO massive access channel described in~\eqref{05}, let $\mathcal{A}= \{1,2, \ldots, k_{n}\}$ denote the active user set in the network, where the total number of users scales as $k_{n}=O(n)$. Assume the codebook size $M_{k}$ for each user is on the same order, the message-length vector with components $\big(R_{1}(n),R_{2}(n),\ldots,R_{k_{n}}(n)\big)$ is asymptotically achievable if
\begin{equation}\label{13}
 R_{k}(n) \leq c_{k}\mathbb{E}_{\mathbf{H}}\Big\{\log\det\big(\mathbf{I}_{N_{R}}+\sum_{t \in
  \mathcal{A}}\mathbf{H}_{t}\mathbf{Q}_{t}\mathbf{H}^{\dagger}_{t}\big)\Big\},
\end{equation}
where $c_{k}=\lim \limits _{n\rightarrow\infty} n\mu_{k}^{(n)}>0$, $\mu_{k}^{(n)} = \frac{\log M_{k}}{\sum_{t \in \mathcal{A}}\log M_{t}} \in (0,1)$.\hfill $\square$

Note that \emph{Theorem 2} has assumed that the codebook sizes $M_{k}$ are on the same order for the users within the network, it then follows that $1/\mu_{k}^{n}=O(n)$ and $\sum_ {k=1}^{k_{n}}c_{k} = n$. Those two conditions may be useful in proving the theorems in later sections.

\emph{Corollary 1}: In the massive access channels, given $k_{n}=O(n)$, the asymptotically individual message-length rate behaves as
\begin{align}\label{14}
  R_{k}(n) \xrightarrow{|\mathcal{A}|\rightarrow\infty}& \, c_{k}N_{R}\log\Big(1+\sum_{t \in \mathcal{A}}\beta_{t}p_{t}\Big)\\
  =&\,c_{k}N_{R}O(\log n)+O(1),
\end{align}
where $p_{k}=\mathrm{Tr}(\mathbf{Q}_{k})$. The term $O(\log n)$ is independent of any other constants but relates only to $n$, and the $O(1)$ constant depends on the pathlosses and transmit powers.\hfill $\square$

The result in~\eqref{14} indicates that due to the effect of channel hardening, the asymptotic rate is close to that when the transmitter has only the statistic knowledge of effective channel gain.

Combining the results in \emph{Theorem 1} and \emph{Theorem 2}, we have the following Theorem for the capacity with massive random access.

\emph{Theorem 3 (Asymmetric Message-Length Capacity for Massive Random Access Channels)}: Let the total number of user be $\ell_{n}$, and assume each user is active with a probability $\alpha_{n}$ such that the average number of active users is $k_{n}=\alpha_{n}\ell_{n}$. Define
\begin{equation}\label{14a}
  \theta_{n}=\frac{\ell_{n} H_{2}(\alpha_{n})}{n\mathbb{E}_{\mathbf{H}}\Big\{\log\det\big(\mathbf{I}_{N_{R}}+
  \sum\limits_{k \in \mathcal{A}_{\mathrm{md}}}\mathbf{H}_{k}\mathbf{Q}_{k}\mathbf{H}^{\dagger}_{k}\big)\Big\}}.
\end{equation}
When $0<\theta_{n}<1$, $k_{n}=O(n)$ and the scaling of $\ell_{n}$ and $k_{n}$ follows~\eqref{09}, the asymmetric message-length capacity for user $k$ is given by
\begin{align}\label{15}
 B_{k}(n)=c_{k}\mathbb{E}_{\mathbf{H}}\Big\{\log\det\big(\mathbf{I}_{N_{R}}+\sum_{t \in
  \mathcal{A}}\mathbf{H}_{t}\mathbf{Q}_{t}\mathbf{H}^{\dagger}_{t}\big)\Big\}-\mu_{k}^{(n)}\ell_{n} H_{2}(\alpha_{n}),
\end{align}
where $c_{k}$ and $\mu_{k}^{(n)}$ are defined in \emph{Theorem 2}.\hfill $\square$

Notice that for each user $k$, rate~\eqref{15} is determined by $c_{k}$ (or $M_{k}$). Since the message set size $M_{k}$ may not necessarily different from each other, given the \emph{Definition 3}, the finite dimension region of MIMO massive access channel is formulated as
\begin{align}\label{16}
  \nonumber &\mathcal{C}_{\mathcal{MAC}}(n)  \triangleq \Bigg\{\big(V_{1}(n),...,V_{J}(n)\big):
  K_{1}V_{1}(n)+\cdots+K_{J} V_{J}(n) \\& \leq n\mathbb{E}_{\mathbf{H}}\Big\{\log\det\big(\mathbf{I}_{N_{R}}+\sum_{k \in \mathcal{A}}\mathbf{H}_{k}\mathbf{Q}_{k}\mathbf{H}^{\dagger}_{k}\big)\Big\}
-\ell_{n} H_{2}(\alpha_{n})\Bigg\},
\end{align}
where $J$ denotes dimension of the region, which is independent of codelength $n$.

\section{Proof of Theorem 1 (UE Identification Cost)}
\subsection{Information Density and the Concentration Inequality}
We begin this section by reviewing the concept of information density~\cite{Polyanskiy2017} and the related concentration inequalities~\cite{Boucheron13}. For the received signal~\eqref{02}, let $\mathbf{S}^{n_{0}}_{\mathcal{A}}=\{\mathbf{S}^{n_{0}}_{\mathcal{A}_{\mathrm{eq}}},\mathbf{S}^{n_{0}}_{\mathcal{A}_{\mathrm{md}}}\}$, where we use $\mathbf{S}^{n_{0}}_{\mathcal{A}}$ to denote the submatrix containing the signals from user set $\mathcal{A}$ only. Given the joint probability $P(\mathbf{Y}^{n_{0}},\mathbf{S}^{n_{0}}_{\mathcal{A}},\mathbf{H}^{n_{0}})$, the conditional information density is defined as
\setcounter{equation}{21}
\begin{equation}\label{22}
  \imath(\mathbf{Y}^{n_{0}};\mathbf{S}^{n_{0}}_{\mathcal{A}_{\mathrm{md}}}|\mathbf{S}^{n_{0}}_{\mathcal{A}_{\mathrm{eq}}},\mathbf{H}^{n_{0}})=
  \log\frac{P(\mathbf{Y}^{n_{0}}|\mathbf{S}^{n_{0}}_{\mathcal{A}_{\mathrm{md}}},\mathbf{S}^{n_{0}}_{\mathcal{A}_{\mathrm{eq}}},\mathbf{H}^{n_{0}})}
  {P(\mathbf{Y}^{n_{0}}|\mathbf{S}^{n_{0}}_{\mathcal{A}_{\mathrm{eq}}},\mathbf{H}^{n_{0}})}.
\end{equation}
Eq.~\eqref{22} is referred to as (conditional) \emph{information density} because when averaging with the joint probability $P(\mathbf{Y}^{n_{0}},\mathbf{S}^{n_{0}}_{\mathcal{A}},\mathbf{H}^{n_{0}})$, we obtain the conditional mutual information, i.e.,
\begin{equation}\label{23}
  \mathbb{E}_{P(\mathbf{Y}^{n_{0}},\mathbf{S}^{n_{0}}_{\mathcal{A}},\mathbf{H}^{n_{0}})}
  \big\{\imath(\mathbf{Y}^{n_{0}};\mathbf{S}^{n_{0}}_{\mathcal{A}_{\mathrm{md}}}|\mathbf{S}^{n_{0}}_{\mathcal{A}_{\mathrm{eq}}},\mathbf{H}^{n_{0}})\big\}
  =\,I(\mathbf{Y}^{n_{0}};\mathbf{S}^{n_{0}}_{\mathcal{A}_{\mathrm{md}}}|\mathbf{S}^{n_{0}}_{\mathcal{A}_{\mathrm{eq}}},\mathbf{H}^{n_{0}}).
\end{equation}

Given the above definition, the following concentration inequality relates to the information density in~\eqref{22} may be useful to prove of the theorems in Section II.

\emph{Lemma 1}: For the arbitrary set partition $\mathcal{A}=\{\mathcal{A}_{\mathrm{eq}},\mathcal{A}_{\mathrm{md}}\}$, let $\mathbf{y}, \mathbf{s}_{\mathcal{A}}$ denote the received and the transmit signals per channel use, respectively. Define $\mathbf{G}_{\mathcal{A}_{\mathrm{md}}}=\sum_{k \in \mathcal{A}_{\mathrm{md}}}\mathbf{H}_{k}\mathbf{Q}_{k}\mathbf{H}^{\dagger}_{k}$, where $\mathbf{H}_{k}$ denotes the random channel matrix for user $k$ in an arbitrary channel use. We then have the following concentration inequality for the signals in~\eqref{02}
\begin{align}\label{24}
  \mathrm{Pr}\Big\{\big|\imath(\mathbf{Y}^{n_{0}};\mathbf{S}^{n_{0}}_{\mathcal{A}_{\mathrm{md}}}|
  \mathbf{S}^{n_{0}}_{\mathcal{A}_{\mathrm{eq}}},\mathbf{H}^{n_{0}})-n_{0}I(\mathbf{y};\mathbf{s}_{\mathcal{A}_{\mathrm{md}}}|\mathbf{s}_{\mathcal{A}_{\mathrm{eq}}},\mathbf{H})\big|
  \geq n_{0}\delta\Big\} \leq  2\exp\Big\{-\frac{n_{0}\delta^{2}}{4c^{2}+2c\delta}\Big\},
\end{align}
where $\delta>0$, $c=32N_{R}+\mathbb{E}_{\mathbf{H}}
\Big[\det\big(\mathbf{I}_{N_{R}}+\mathbf{G}_{\mathcal{A}_{\mathrm{md}}}\big)\Big]
e^{-I(\mathbf{y};\mathbf{s}_{\mathcal{A}_{\mathrm{md}}}|\mathbf{s}_{\mathcal{A}_{\mathrm{eq}}},\mathbf{H})}$ is some constant, and the mutual information is given by
\begin{equation}\label{24a}
I(\mathbf{y};\mathbf{s}_{\mathcal{A}_{\mathrm{md}}}|\mathbf{s}_{\mathcal{A}_{\mathrm{eq}}},\mathbf{H})
=\mathbb{E}_{\mathbf{H}}\Big\{\log\det\Big(\mathbf{I}_{N_{R}}+\sum_{k \in \mathcal{A}_{\mathrm{md}}}\mathbf{H}_{k}\mathbf{Q}_{k}\mathbf{H}^{\dagger}_{k}\Big)\Big\}.
\end{equation}
\hfill $\square$

\emph{Proof}: See Appendix A.

\subsection{Proof of Achievability}
Consider a \emph{maximum likelihood} (ML) decoder for the active UE detection. Mathematically, the ML decoder is assumed to search an arbitrary subset $\tilde{\mathcal{A}} \subseteq \mathcal{T}$ with such that
\begin{equation}\label{25}
  \mathcal{A} \equiv \hat{\mathcal{A}} =  \arg \max_{\tilde{\mathcal{A}} \subseteq \mathcal{T}} P(\mathbf{Y}^{n_{0}}|\mathbf{S}^{n_{0}}_{\tilde{\mathcal{A}}},\mathbf{H}^{n_{0}}).
\end{equation}
We define the error event
\begin{align}\label{26}
  \nonumber \Xi_{i}=\big\{ \exists \, \hat{\mathcal{A}} \neq \mathcal{A}:
  P(\mathbf{Y}^{n_{0}}|\mathbf{S}^{n_{0}}_{\hat{\mathcal{A}}},&\,\mathbf{H}^{n_{0}})>P(\mathbf{Y}^{n_{0}}|\mathbf{S}^{n_{0}}_{\mathcal{A}},
  \mathbf{H}^{n_{0}}),\\
  &|\mathcal{A}_{\mathrm{fa}}|=|\mathcal{A}_{\mathrm{md}}|=i, |\hat{\mathcal{A}}|=|\mathcal{A}|=k_{\ell}\big\},
\end{align}
for all $i\in \{1,\ldots,k_{\ell}\}$, as there exists a $k_{\ell}$-size set $\hat{\mathcal{A}}$ that has resulted in $i$ false alarm users and $i$ misdetection users, and is deemed to be more likely by the decoder. The relationship between sets $\hat{\mathcal{A}}$ and $\mathcal{A}$ is illustrated in Fig.~\ref{Fig.2}.

Define the set $\Psi_{i}$ as
\begin{equation}\label{27}
  \Psi_{i} = \big\{\hat{\mathcal{A}} \subseteq \mathcal{T}: |\mathcal{A}_{\mathrm{fa}}|=i, |\mathcal{A}_{\mathrm{eq}}|=k_{\ell}-i\big\},
\end{equation}
i.e., the collection of decoded sets $\hat{\mathcal{A}}$ that result in $i$ false alarm users and $i$ misdetection users. The conditional probability follows
\begin{equation}\label{28}
  \mathrm{Pr}\big\{\Xi_{i}|\mathbf{S}^{n_{0}}_{\mathcal{A}},\mathbf{H}^{n_{0}},\mathbf{Y}^{n_{0}}\big\}\leq
  \mathrm{Pr}\Big(\bigcup_{\hat{\mathcal{A}} \in \Psi_{i}}\Xi_{i}(\hat{\mathcal{A}})\mid \mathbf{S}^{n_{0}}_{\mathcal{A}},\mathbf{H}^{n_{0}},\mathbf{Y}^{n_{0}} \Big).
\end{equation}
For the particular decoded set $\hat{\mathcal{A}}$, the conditional error probability $P(\Xi_{i}(\hat{\mathcal{A}})\mid \mathbf{S}^{n_{0}}_{\mathcal{A}},\mathbf{H}^{n_{0}},\mathbf{Y}^{n_{0}})$ is calculated as
\begin{align}
  \label{29a}\mathrm{Pr}\big(\Xi_{i}(\hat{\mathcal{A}})\mid \mathbf{S}^{n_{0}}_{\mathcal{A}},\mathbf{H}^{n_{0}},\mathbf{Y}^{n_{0}} \big) &= \sum_{\mathbf{S}^{n_{0}}_{\hat{\mathcal{A}}}: \,
  P(\mathbf{Y}^{n_{0}}|\mathbf{S}^{n_{0}}_{\hat{\mathcal{A}}},\mathbf{H}^{n_{0}})>
  P(\mathbf{Y}^{n_{0}}|\mathbf{S}^{n_{0}}_{\mathcal{A}},\mathbf{H}^{n_{0}})} Q(\mathbf{S}^{n_{0}}_{\hat{\mathcal{A}}}|\mathbf{S}^{n_{0}}_{\mathcal{A}}) \\
  \label{29b}& \leq \sum_{\mathbf{S}^{n_{0}}_{\mathcal{A}_{\mathrm{fa}}}}  Q(\mathbf{S}^{n_{0}}_{\mathcal{A}_{\mathrm{fa}}}) \frac{P(\mathbf{Y}^{n_{0}}|\mathbf{S}^{n_{0}}_{\hat{\mathcal{A}}},\mathbf{H}^{n_{0}})^{s}}
  {P(\mathbf{Y}^{n_{0}}|\mathbf{S}^{n_{0}}_{\mathcal{A}},\mathbf{H}^{n_{0}})^{s}},
\end{align}
for any $s>0$, where~\eqref{29b} is due to the conditional on $\mathcal{A}$ and the set partition $\hat{\mathcal{A}}=\{\mathcal{A}_{\mathrm{eq}}, \mathcal{A}_{\mathrm{fa}}\}$.

By the union bound and Gallager's $\rho$ trick~\cite{Gallager68},~\eqref{28} is upper bounded as
\begin{align}
  \label{30a}\mathrm{Pr}\big\{\Xi_{i}|\mathbf{S}^{n_{0}}_{\mathcal{A}},\mathbf{H}^{n_{0}},\mathbf{Y}^{n_{0}}\big\}
  &\leq\sum_{\hat{\mathcal{A}} \in \Psi_{i}}\mathrm{Pr}\big(\Xi_{i}(\hat{\mathcal{A}})\mid
  \mathbf{S}_{\mathcal{A}}^{n_{0}}, \mathbf{H}^{n_{0}}, \mathbf{Y}^{n_{0}} \big)\\
  \label{30b} &\leq \sum_{\mathcal{A}_{\mathrm{eq}}\in \Psi_{i}}\Bigg(\sum_{\mathcal{A}_{\mathrm{fa}}\in \Psi_{i}}\mathrm{Pr}\big(\Xi_{i}(\hat{\mathcal{A}})\mid \mathbf{S}^{n_{0}}_{\mathcal{A}},\mathbf{H}^{n_{0}},\mathbf{Y}^{n_{0}} \big)\Bigg)^{\rho},
\end{align}
for some $\rho \in [0,1]$.
Since the number of sets $\mathcal{A}_{\mathrm{eq}}$ is no greater than $\binom{k_{\ell}}{i}$, while the number of sets $\mathcal{A}_{\mathrm{fa}}$ is no greater than $\binom{\ell-k_{\ell}}{i}$, plugging~\eqref{29b} into~\eqref{30b} we have
\begin{align}\label{31}
  \mathrm{Pr}\big\{\Xi_{i}|\mathbf{S}^{n_{0}}_{\mathcal{A}},\mathbf{H}^{n_{0}},\mathbf{Y}^{n_{0}}\big\}
  \leq  \max_{\mathcal{A}_{\mathrm{eq}} \subseteq \Psi_{i}}
  \exp\Bigg\{-\Bigg[E_{o}(\rho,s,\mathcal{A}_{\mathrm{eq}})-\rho\log\binom{\ell-k_{\ell}}{i}
  -\log\binom{k_{\ell}}{i}\Bigg]\Bigg\},
\end{align}
where the exponent is given by
\begin{equation}\label{32}
  E_{o}(\rho,s,\mathcal{A}_{\mathrm{eq}}) = \rho \log\frac{P(\mathbf{Y}^{n_{0}}|\mathbf{S}^{n_{0}}_{\mathcal{A}},\mathbf{H}^{n_{0}})^{s}}
  {\sum_{\mathbf{S}^{n_{0}}_{\mathcal{A}_{\mathrm{fa}}}}Q(\mathbf{S}^{n_{0}}_{\mathcal{A}_{\mathrm{fa}}})
  P(\mathbf{Y}^{n_{0}}|\mathbf{S}^{n_{0}}_{\hat{\mathcal{A}}},\mathbf{H}^{n_{0}})^{s}}.
\end{equation}

In order to prove that the error probability for ML decoder is vanishing, it is equivalent to show that
\begin{equation}\label{33}
  E_{o}(\rho,s,\mathcal{A}_{\mathrm{eq}})-\rho\log\binom{\ell-k_{\ell}}{i}-\log\binom{k_{\ell}}{i}\rightarrow\infty
\end{equation}
for the arbitrary set partitions $\mathcal{A}=\{\mathcal{A}_{\mathrm{md}},\mathcal{A}_{\mathrm{eq}}\}$ as $\ell$ goes to infinity. This was equivalent to say the following inequality
\begin{align}\label{34}
  \mathrm{Pr}\Bigg\{\bigcup_{\mathcal{A}_{\mathrm{md}},\mathcal{A}_{\mathrm{eq}}}\Big[E_{o}(\rho,s,\mathcal{A}_{\mathrm{eq}})
  -\rho\log\binom{\ell-k_{\ell}}{i}-\log\binom{k_{\ell}}{i}\leq\gamma\Big]\Bigg\} \leq \varphi_{\mathcal{A}_{\mathrm{md}}}(\cdot),
\end{align}
holds for some finite constant $\gamma$ that is independent of $\ell$, and $\varphi_{\mathcal{A}_{\mathrm{md}}}(\cdot)$ is some vanishing functions, i.e., $\varphi_{\mathcal{A}_{\mathrm{md}}}(\cdot)\rightarrow0$ as $\ell\rightarrow\infty$, where $\mathrm{Pr}(\cdot)$ is with respect to (w.r.t.) the joint probability $P(\mathbf{Y}^{n_{0}}|\mathbf{S}^{n_{0}}_{\mathcal{A}},\mathbf{H}^{n_{0}})
Q(\mathbf{S}^{n_{0}}_{\mathcal{A}})Q(\mathbf{H}^{n_{0}})$. The union in~\eqref{34} follows because
\begin{equation}\label{35}
 \mathrm{Pr}\Big(\max_{k} X_{k}\leq Y\Big) = \mathrm{Pr}\Big(\bigcup_{k}\big\{X_{k}\leq Y\big\}\Big)
\end{equation}
for the finite number of random variables $X_{k}$.

Let $\zeta_{i}=\rho\log\binom{\ell-k_{\ell}}{i}+\log\binom{k_{\ell}}{i}+\gamma$, and define
\begin{equation}\label{35b}
  E_{o}'(\rho,s,\mathcal{A}_{\mathrm{eq}})=\rho\log\frac{\sum_{\mathbf{S}^{n_{0}}_{\mathcal{A}_{\mathrm{fa}}}}Q(\mathbf{S}^{n_{0}}_{\mathcal{A}_{\mathrm{fa}}})
  P(\mathbf{Y}^{n_{0}}|\mathbf{S}^{n_{0}}_{\hat{\mathcal{A}}},\mathbf{H}^{n_{0}})^{s}}{P(\mathbf{Y}^{n_{0}}|\mathbf{S}^{n_{0}}_{\mathcal{A}_{\mathrm{eq}}},\mathbf{H}^{n_{0}})^{s}}.
\end{equation}
The L.H.S. of inequality~\eqref{34} is shown to be
\begin{align}
  \nonumber &\mathrm{Pr}\Bigg\{\bigcup_{\mathcal{A}_{\mathrm{md}},\mathcal{A}_{\mathrm{eq}}}\Big[E_{o}(\rho,s,\mathcal{A}_{\mathrm{eq}})\leq\zeta_{i}\Big]\Bigg\}\\
  \nonumber\leq&\, \mathrm{Pr}\Bigg\{\bigcup_{\mathcal{A}_{\mathrm{md}},\mathcal{A}_{\mathrm{eq}}}\Big[E_{o}(\rho,s,\mathcal{A}_{\mathrm{eq}})\leq\zeta_{i} \cap E_{o}'(\rho,s,\mathcal{A}_{\mathrm{eq}})\leq\zeta_{i}'\Big]\Bigg\}\\
  \label{36a}&+\mathrm{Pr}\Bigg\{\bigcup_{\mathcal{A}_{\mathrm{md}},\mathcal{A}_{\mathrm{eq}}}\Big[E_{o}'(\rho,s,\mathcal{A}_{\mathrm{eq}})>\zeta_{i}'\Big]\Bigg\}\\
  \nonumber\leq&\, \mathrm{Pr}\Bigg\{\bigcup_{\mathcal{A}_{\mathrm{md}},\mathcal{A}_{\mathrm{eq}}}
  \Big[s\rho\cdot\imath(\mathbf{Y}^{n_{0}};\mathbf{S}^{n_{0}}_{\mathcal{A}_{\mathrm{md}}}|\mathbf{S}^{n_{0}}_{\mathcal{A}_{\mathrm{eq}}},\mathbf{H}^{n_{0}})
  \leq\zeta_{i}+\zeta_{i}'\Big]\Bigg\}\\
  \label{36b}&+\mathrm{Pr}\Bigg\{\bigcup_{\mathcal{A}_{\mathrm{md}},\mathcal{A}_{\mathrm{eq}}}\Big[E_{o}'(\rho,s,\mathcal{A}_{\mathrm{eq}})>\zeta_{i}'\Big]\Bigg\}
\end{align}
for some $\zeta_{i}'$, where~\eqref{36a} is by the relation
\begin{align}\label{37}
  \nonumber \mathrm{Pr}(A)&=\mathrm{Pr}(A)\big(\mathrm{Pr}(B)+\mathrm{Pr}(\bar{B})\big)\\
  &\leq \mathrm{Pr}(A\cap B)+ \mathrm{Pr}(\bar{B})
\end{align}
for the independent events $A$ and $B$.

The second term in~\eqref{36b} is given by
\begin{align}
  \nonumber &\mathrm{Pr}\Bigg\{\bigcup_{\mathcal{A}_{\mathrm{md}},\mathcal{A}_{\mathrm{eq}}}\Big[E_{o}'(\rho,s,\mathcal{A}_{\mathrm{eq}}) > \zeta_{i}'\Big]\Bigg\}\\
  \nonumber =&\sum_{\mathcal{A}_{\mathrm{md}},\mathcal{A}_{\mathrm{eq}}}
  \sum_{\mathbf{Y}^{n_{0}},\mathbf{S}^{n_{0}}_{\mathcal{A}_{\mathrm{eq}}},\mathbf{H}^{n_{0}}}
  P(\mathbf{Y}^{n_{0}}|\mathbf{S}^{n_{0}}_{\mathcal{A}_{\mathrm{eq}}},\mathbf{H}^{n_{0}})Q(\mathbf{S}^{n_{0}}_{\mathcal{A}_{\mathrm{eq}}})Q(\mathbf{H}^{n_{0}})\\
  \label{38a}&\qquad \times
  1\Big\{\rho\log\frac{\sum_{\mathbf{S}^{n_{0}}_{\mathcal{A}_{\mathrm{fa}}}}Q(\mathbf{S}^{n_{0}}_{\mathcal{A}_{\mathrm{fa}}})
  P(\mathbf{Y}^{n_{0}}|\mathbf{S}^{n_{0}}_{\hat{\mathcal{A}}},\mathbf{H}^{n_{0}})^{s}}{P(\mathbf{Y}^{n_{0}}|\mathbf{S}^{n_{0}}_{\mathcal{A}_{\mathrm{eq}}},\mathbf{H}^{n_{0}})^{s}} > \zeta_{i}'\Big\}\\
  \label{38b}\leq& \,\sum_{i=1}^{k_{\ell}}\binom{k_{\ell}}{i}e^{-\frac{\zeta_{i}'}{\rho}}
  \cdot\sum_{\mathbf{Y}^{n_{0}},\mathbf{S}^{n_{0}}_{\mathcal{A}_{\mathrm{eq}}},\mathbf{H}^{n_{0}}}
  Q(\mathbf{S}^{n_{0}}_{\mathcal{A}_{\mathrm{eq}}},\mathbf{H}^{n_{0}})\Big(\sum_{\mathbf{S}^{n_{0}}_{\mathcal{A}_{\mathrm{fa}}}}Q(\mathbf{S}^{n_{0}}_{\mathcal{A}_{\mathrm{fa}}})
  P(\mathbf{Y}^{n_{0}}|\mathbf{S}^{n_{0}}_{\hat{\mathcal{A}}},\mathbf{H}^{n_{0}})^{s}\Big)^{1/s}\\
  \label{38c}\leq& \,\sum_{i=1}^{k_{\ell}}\binom{k_{\ell}}{i}e^{-\frac{\zeta_{i}'}{\rho}},
\end{align}
where~\eqref{38c} is due to the fact that by choosing $s=\frac{1}{1+\rho}$, the following inequality always holds~\cite{Gallager68}
\begin{equation}\label{39}
 -\log \sum_{\mathbf{Y}^{n_{0}},\mathbf{S}^{n_{0}}_{\mathcal{A}_{\mathrm{eq}}},\mathbf{H}^{n_{0}}}
  \Big(\sum_{\mathbf{S}^{n_{0}}_{\mathcal{A}_{\mathrm{fa}}}}Q(\mathbf{S}^{n_{0}}_{\mathcal{A}_{\mathrm{fa}}})
  P(\mathbf{Y}^{n_{0}},\mathbf{S}^{n_{0}}_{\mathcal{A}_{\mathrm{eq}}},\mathbf{H}^{n_{0}}
  |\mathbf{S}^{n_{0}}_{\mathcal{A}_{\mathrm{fa}}})^{\frac{1}{1+\rho}}\Big)^{1+\rho}\geq0.
\end{equation}
By choosing $\zeta_{i}'=\rho\log\big[\frac{k_{\ell}}{\delta_{1}}\binom{k_{\ell}}{i}\big]$ for some vanishing factor $\delta_{1} \in (0,1)$, we obtain
\begin{align}\label{40}
  \mathrm{Pr}\Bigg\{&\bigcup_{\mathcal{A}_{\mathrm{md}},\mathcal{A}_{\mathrm{eq}}}\Big[\rho\log\frac{\sum_{\mathbf{S}^{n_{0}}_{\mathcal{A}_{\mathrm{fa}}}}Q(\mathbf{S}^{n_{0}}_{\mathcal{A}_{\mathrm{fa}}})
  P(\mathbf{Y}^{n_{0}}|\mathbf{S}^{n_{0}}_{\hat{\mathcal{A}}},\mathbf{H}^{n_{0}})^{s}}
  {P(\mathbf{Y}^{n_{0}}|\mathbf{S}^{n_{0}}_{\mathcal{A}_{\mathrm{eq}}},\mathbf{H}^{n_{0}})^{s}}
  > \zeta_{i}'\Big]\Bigg\}\leq\delta_{1}.
\end{align}

We now left with the first term in~\eqref{36b}. Given $\zeta_{i}'=\rho\log\big[\frac{k_{\ell}}{\delta}\binom{k_{\ell}}{i}\big]$ and $s=\frac{1}{1+\rho}$, the first term in~\eqref{36b} can be written as
\begin{align}\label{41}
  \nonumber \mathrm{Pr}\Bigg\{\bigcup_{\mathcal{A}_{\mathrm{md}},\mathcal{A}_{\mathrm{eq}}}\Big[\frac{\rho}{1+\rho} \imath(\mathbf{Y}^{n_{0}};\mathbf{S}^{n_{0}}_{\mathcal{A}_{\mathrm{md}}}&|\mathbf{S}^{n_{0}}_{\mathcal{A}_{\mathrm{eq}}},\mathbf{H}^{n_{0}}) \leq \rho\log\binom{\ell-k_{\ell}}{i}\\
 &+(\rho+1)\log\binom{k_{\ell}}{i}+\rho\log\frac{k_{\ell}}{\delta_{1}}+\gamma\Big]\Bigg\}.
\end{align}
To prove the probability~\eqref{41} is vanishing as $\ell\rightarrow\infty$, we make the following assumption for $n_{0}$
\begin{align}\label{42}
  \nonumber\rho\log\binom{\ell-k_{\ell}}{i}+(\rho+1)&\log\binom{k_{\ell}}{i}+
  \rho\log\frac{k_{\ell}}{\delta_{1}}+\gamma \\
  \leq & \, \frac{\rho(1-\delta_{2})}{1+\rho} n_{0}I(\mathbf{y};\mathbf{s}_{\mathcal{A}_{\mathrm{md}}}|\mathbf{s}_{\mathcal{A}_{\mathrm{eq}}},\mathbf{H}),
\end{align}
where $\delta_{2} \in (0,1)$ is a small enough constant.

By \emph{Lemma 1} and setting $\delta=\delta_{2}I(\mathbf{y};\mathbf{s}_{\mathcal{A}_{\mathrm{md}}}|\mathbf{s}_{\mathcal{A}_{\mathrm{eq}}},\mathbf{H})$ for some $\delta_{2}\in(0,1)$, we have the following inequality,
\begin{align}\label{43}
  \nonumber \mathrm{Pr}\Big\{\imath(&\mathbf{Y}^{n_{0}};\mathbf{S}^{n_{0}}_{\mathcal{A}_{\mathrm{md}}}
  |\mathbf{S}^{n_{0}}_{\mathcal{A}_{\mathrm{eq}}},\mathbf{H}^{n_{0}})\leq (1-\delta_{2})n_{0}I(\mathbf{y};\mathbf{s}_{\mathcal{A}_{\mathrm{md}}}|\mathbf{s}_{\mathcal{A}_{\mathrm{eq}}},\mathbf{H})\Big\}\\
  &\leq 2\exp\Bigg\{-\frac{n_{0}\big[\delta_{2}
  I(\mathbf{y};\mathbf{s}_{\mathcal{A}_{\mathrm{md}}}|\mathbf{s}_{\mathcal{A}_{\mathrm{eq}}},\mathbf{H})\big]^{2}}
  {4c^{2}+2c\delta_{2}I(\mathbf{y};\mathbf{s}_{\mathcal{A}_{\mathrm{md}}}|\mathbf{s}_{\mathcal{A}_{\mathrm{eq}}},\mathbf{H})}\Bigg\},
\end{align}
for a given set partition $\mathcal{A}=\{\mathcal{A}_{\mathrm{md}},\mathcal{A}_{\mathrm{eq}}\}$. By the union bound, and combining~\eqref{41},~\eqref{42} with~\eqref{43}, we have
\begin{align}\label{44}
  \nonumber \mathrm{Pr}\Bigg\{\bigcup_{\mathcal{A}_{\mathrm{md}},\mathcal{A}_{\mathrm{eq}}}\Big[&\frac{\rho}{1+\rho}
  \imath(\mathbf{Y}^{n_{0}};\mathbf{S}^{n_{0}}_{\mathcal{A}_{\mathrm{md}}}|\mathbf{S}^{n_{0}}_{\mathcal{A}_{\mathrm{eq}}},\mathbf{H}^{n_{0}})
  \leq \rho\log\binom{\ell-k_{\ell}}{i} \\
  \nonumber &+(\rho+1)\log\binom{k_{\ell}}{i}+
  \rho\log\frac{k_{\ell}}{\delta_{1}}+\gamma\Big]\Bigg\}\\
  \leq&\sum_{i=1}^{k_{\ell}}\binom{k_{\ell}}{i}\psi_{i}(n_{0},\delta_{2}),
\end{align}
where $\psi_{i}(n_{0},\delta_{2})$ is given by
\begin{equation}\label{45}
  \psi_{i}(n_{0},\delta_{2})=2\exp\Bigg\{-\frac{n_{0}\big[\delta_{2}
  I(\mathbf{y};\mathbf{s}_{\mathcal{A}_{\mathrm{md}}}|\mathbf{s}_{\mathcal{A}_{\mathrm{eq}}},\mathbf{H})\big]^{2}}
  {4c^{2}+2c\delta_{2}I(\mathbf{y};\mathbf{s}_{\mathcal{A}_{\mathrm{md}}}|\mathbf{s}_{\mathcal{A}_{\mathrm{eq}}},\mathbf{H})}\Bigg\}.
\end{equation}

We now investigated the conditions for $n_{0}$ such that the probability~\eqref{44} vanishes. Note that the upper bound of is given by
\begin{align}\label{46}
  \sum_{i=1}^{k_{\ell}}\binom{k_{\ell}}{i}\psi_{i}(n_{0},\delta_{2})
  \leq \sum_{i=1}^{k_{\ell}} 2\exp\Bigg\{\log\binom{k_{\ell}}{i}-\frac{n_{0}\big[\delta_{2}
  I(\mathbf{y};\mathbf{s}_{\mathcal{A}_{\mathrm{md}}}|\mathbf{s}_{\mathcal{A}_{\mathrm{eq}}},\mathbf{H})\big]^{2}}
  {4c^{2}+2c\delta_{2}I(\mathbf{y};\mathbf{s}_{\mathcal{A}_{\mathrm{md}}}|\mathbf{s}_{\mathcal{A}_{\mathrm{eq}}},\mathbf{H})}\Bigg\}.
\end{align}
Thus, proving of~\eqref{46} vanishes is equivalent to show
\begin{align}\label{47}
  -\log\binom{k_{\ell}}{i}-\log k_{\ell}+\frac{n_{0}\big[\delta_{2}
  I(\mathbf{y};\mathbf{s}_{\mathcal{A}_{\mathrm{md}}}|\mathbf{s}_{\mathcal{A}_{\mathrm{eq}}},\mathbf{H})\big]^{2}}
  {4c^{2}+2c\delta_{2}I(\mathbf{y};\mathbf{s}_{\mathcal{A}_{\mathrm{md}}}|\mathbf{s}_{\mathcal{A}_{\mathrm{eq}}},\mathbf{H})}\rightarrow\infty
\end{align}
for any $i=1,2,\ldots,k_{\ell}$ as $\ell$ goes to infinity, where $c$ is some constant.
By choosing $\rho=\epsilon'$ for some $\epsilon' \in (0,1)$,~\eqref{42} further assumes the following condition for $n_{0}$
\begin{equation}\label{48}
  n_{0}\geq (1+\epsilon')\frac{\log\binom{\ell-k_{\ell}}{i}+\frac{1+\epsilon'}{\epsilon'}\log\binom{k_{\ell}}{i}
  +\log\frac{k_{\ell}}{\delta_{1}}+\frac{\gamma}{\epsilon'}}{(1-\delta_{2})
  I(\mathbf{y};\mathbf{s}_{\mathcal{A}_{\mathrm{md}}}|\mathbf{s}_{\mathcal{A}_{\mathrm{eq}}},\mathbf{H})},
\end{equation}
for $\forall i\in [1,k_{\ell}]$.

\emph{Proposition 1}: Let $i=|\mathcal{A}_{\mathrm{md}}|$, $\ell$ and $k_{\ell}$ denote the total number of users and the number of active users on average, which satisfy the relationship~\eqref{09}. Given the signature length $n_{0}$ in~\eqref{48}, the condition~\eqref{47} holds asymptotically as $\ell\rightarrow\infty$. Further, by the assumption $\ell\gg k_{\ell}$, Eq.~\eqref{11} and~\eqref{48} are equivalent asymptotically. Therefore, by choosing the signature length~\eqref{11}, the error probability~\eqref{28} can be made arbitrary small as $\ell\rightarrow\infty$.\hfill $\square$

\emph{Proof}: see Appendix B.

Given the \emph{Proposition 1}, averaging~\eqref{28} with the joint probability $P(\mathbf{Y},\mathbf{\Phi}_{\mathcal{A}})$, we have $\mathrm{Pr}(\Xi_{i})\rightarrow0, \forall i\in [1,k_{\ell}]$.

\subsection{Proof of the Converse}
The proof of converse is based on Theorem 2 in~\cite{Scarlett17}. The idea is that a genie can reveal the set $\mathcal{A}_{\mathrm{eq}}$ to decoder, and based on that the decoder outputs the estimate for $\mathcal{A}_{\mathrm{md}}$. For the clarity of the proof, we rewrite the theorem, which states the lower bound of error probability as follows
\begin{align}\label{49}
  \mathrm{Pr}\big\{\Xi_{i}|\mathbf{S}^{n_{0}}_{\mathcal{A}},\mathbf{H}^{n_{0}},\mathbf{Y}^{n_{0}}\big\} \geq
  \mathrm{Pr}\Big\{\imath(\mathbf{Y}^{n_{0}};\mathbf{S}^{n_{0}}_{\mathcal{A}_{\mathrm{md}}}|\mathbf{S}^{n_{0}}_{\mathcal{A}_{\mathrm{eq}}},\mathbf{H}^{n_{0}})\leq \log\binom{\ell-k_{\ell}+i}{i}+\log\delta_{1}\Big\}-\delta_{1},
\end{align}
for some fix $\delta_{1}>0$, $i=|\mathcal{A}_{\mathrm{md}}|$.

To find the necessary condition, we first assume the following inequalities for $n_{0}$
\begin{equation}\label{50}
  \log\binom{\ell-k_{\ell}+i}{i}+\log\delta_{1}\geq n_{0}(1+\delta_{2})
  I(\mathbf{y};\mathbf{s}_{\mathcal{A}_{\mathrm{md}}}|\mathbf{s}_{\mathcal{A}_{\mathrm{eq}}},\mathbf{H}),
\end{equation}
and
\begin{align}\label{51}
  \mathrm{Pr}\Big\{\imath(\mathbf{Y}^{n_{0}};\mathbf{S}^{n_{0}}_{\mathcal{A}_{\mathrm{md}}}|\mathbf{S}^{n_{0}}_{\mathcal{A}_{\mathrm{eq}}},\mathbf{H}^{n_{0}})
  \leq n_{0}(1+\delta_{2})
  I(\mathbf{y};\mathbf{s}_{\mathcal{A}_{\mathrm{md}}}|\mathbf{s}_{\mathcal{A}_{\mathrm{eq}}},\mathbf{H})\Big\}
  \geq 1-\varphi_{i}(\cdot),
\end{align}
for some functions $\varphi_{i}(\cdot)$. Combining~\eqref{50} and~\eqref{51}, the error probability $\mathrm{Pr}\big\{\Xi_{i}|\mathbf{S}^{n_{0}}_{\mathcal{A}},\mathbf{H}^{n_{0}},\mathbf{Y}^{n_{0}}\big\}$ is lower bounded by $1-\varphi_{i}(\cdot)$. By \emph{Lemma 1}  and setting $\delta=\delta_{2}I(\mathbf{y};\mathbf{s}_{\mathcal{A}_{\mathrm{md}}}|\mathbf{s}_{\mathcal{A}_{\mathrm{eq}}},\mathbf{H})$ for some $\delta_{2}\in(0,1)$, the tail bound of~\eqref{51} is given by
\begin{equation}\label{52}
  1-2\exp\Bigg\{-\frac{n_{0}\big[\delta_{2}
  I(\mathbf{y};\mathbf{s}_{\mathcal{A}_{\mathrm{md}}}|\mathbf{s}_{\mathcal{A}_{\mathrm{eq}}},\mathbf{H})\big]^{2}}
  {4c^{2}+2c\delta_{2}I(\mathbf{y};\mathbf{s}_{\mathcal{A}_{\mathrm{md}}}|\mathbf{s}_{\mathcal{A}_{\mathrm{eq}}},\mathbf{H})}\Bigg\}.
\end{equation}
In addition,~\eqref{50} implies that
\begin{equation}\label{53}
  n_{0} \leq \frac{\log\binom{\ell-k_{\ell}+i}{i}+\log\delta_{1}}{(1+\delta_{2})
  I(\mathbf{y};\mathbf{s}_{\mathcal{A}_{\mathrm{md}}}|\mathbf{s}_{\mathcal{A}_{\mathrm{eq}}},\mathbf{H})}.
\end{equation}

By choosing $\delta_{1}\rightarrow 0$ sufficient slowly in~\eqref{53}, and note that the partition for set $\mathcal{A}=\{\mathcal{A}_{\mathrm{md}},\mathcal{A}_{\mathrm{eq}}\}$ is arbitrary, we obtain the converse result. To show~\eqref{12} renders the error probability $\mathrm{Pr}\big\{\Xi_{i}|\mathbf{S}^{n_{0}}_{\mathcal{A}},\mathbf{H}^{n_{0}},\mathbf{Y}^{n_{0}}\big\}
\rightarrow1$, consider the worst case for $i=1$ in~\eqref{12}, in which $n_{0}$ behaves as $\Omega(\log(\ell-k_{\ell}))$. Substitute that into~\eqref{52}, and note that $\varphi_{i}(\cdot)$ is on the order of $O(e^{-n_{0}})$, the tail bound thus approaches to $1$ as $\ell$ goes to infinity. For the above analysis, the equality is assumed in~\eqref{12} since the decoder can do no better with the less signature length. Averaging~\eqref{28} with the joint probability $P(\mathbf{S}^{n_{0}}_{\mathcal{A}},\mathbf{H}^{n_{0}},\mathbf{Y}^{n_{0}})$, we have $\mathrm{Pr}(\Xi_{i})\rightarrow1, \forall i\in [1,k_{\ell}]$.

\section{Proof of Theorem 2 (Asymmetric Capacity)}
After training phase, the BS may know exactly the set of active UEs within the network. In this section, we prove \emph{Theorem 2}, which characterizes the asymmetric capacity of MIMO massive access channel when set $\mathcal{A}$ is known perfectly by the receiver.
\subsection{Proof of Achievability}
Let $\mathcal{A}_{l}$ denote the $l^{th}$ subset of $\mathcal{A}$ with size $\tilde{k}$. Following an analogous way in deriving Eq.~\eqref{25}-\eqref{32}, we have the upper bound of the decoding error probability that $\mathcal{A}_{l} \subseteq \mathcal{A}$ users incur a decoding error,
\begin{align}\label{54}
  \mathrm{Pr}\big\{\check{\Xi}_{\tilde{k}}|\check{\mathbf{S}}^{n}_{\mathcal{A}},\mathbf{H}^{n},\check{\mathbf{Y}}^{n}\big\}\leq
  \max_{\mathcal{A}_{l} \subseteq \Psi_{\tilde{k}}}
  \exp\Bigg\{-\Bigg[e_{o}(\rho,s,\mathcal{A}_{l}&)-\rho\sum_{k\in\mathcal{A}_{l}}R_{k}(n)-\log\binom{k_{n}}{\tilde{k}}\Bigg]\Bigg\},
\end{align}
where $R_{k}(n)=\log M_{k}$, $\check{\Xi}_{\tilde{k}}$ is the event that users in $\mathcal{A}_{l} \subseteq \mathcal{A}$ incur a decoding error, and
\begin{equation}\label{55}
  \Psi_{\tilde{k}} = \Big\{\mathcal{A}_{l}:\mathcal{A}_{l} \subseteq \mathcal{A}, |\mathcal{A}_{l}|=\tilde{k}, l \in \Big[1,\binom{k_{n}}{\tilde{k}}\Big]\Big\},
\end{equation}
and
\begin{equation}\label{56}
  e_{o}(\rho,s,\mathcal{A}_{l})= \rho \log\frac{P(\check{\mathbf{Y}}^{n}|\check{\mathbf{S}}^{n}_{\mathcal{A}},\mathbf{H}^{n})^{s}}
  {\sum_{\check{\mathbf{S}}^{n^{'}}_{\mathcal{A}_{l}}}Q(\check{\mathbf{S}}^{n^{'}}_{\mathcal{A}_{l}})
  P(\check{\mathbf{Y}}^{n}|\check{\mathbf{S}}^{n^{'}}_{\mathcal{A}},\mathbf{H}^{n})^{s}}.
\end{equation}
Averaging~\eqref{54} with the joint probability $P(\check{\mathbf{S}}^{n}_{\mathcal{A}},\mathbf{H}^{n},\check{\mathbf{Y}}^{n})$ and note that $\check{\mathbf{S}}^{n^{'}}_{\mathcal{A}_{l}}$ is the dummy variables, we have
\begin{align}\label{57}
  \mathrm{Pr}\big\{\check{\Xi}_{\tilde{k}}\big\}\leq
  \max_{\mathcal{A}_{l} \subseteq \Psi_{\tilde{k}}}
 \exp\Bigg\{-\Bigg[E_{o}(\rho,s,\mathcal{A}_{l})-\rho\sum_{k\in\mathcal{A}_{l}}R_{k}(n)
 -\log\binom{k_{n}}{\tilde{k}}\Bigg]\Bigg\},
\end{align}
where let $s=1/(1+\rho)$, and $\mathcal{A}_{l}^{c}$ denote the complementary set of $\mathcal{A}_{l}$, the error exponent $E_{o}(\rho,\mathcal{A}_{l})$ is given by
\begin{align}\label{58}
   E_{o}(\rho,\mathcal{A}_{l})=-\log\mathbb{E}_{\mathbf{H}^{n}}
  \Bigg\{\int Q(\check{\mathbf{S}}^{n}_{\mathcal{A}_{l}^{c}})
  \Bigg[\int Q(\check{\mathbf{S}}^{n}_{\mathcal{A}_{l}})
  P(\check{\mathbf{Y}}^{n}|\check{\mathbf{S}}^{n}_{\mathcal{A}},\mathbf{H}^{n})^{\frac{1}{1+\rho}}d\check{\mathbf{S}}^{n}_{\mathcal{A}_{l}} \Bigg]^{1+\rho} d\check{\mathbf{S}}^{n}_{\mathcal{A}_{l}^{c}}d\check{\mathbf{Y}}^{n}\Bigg\}.
\end{align}
For the random codewords $\check{\mathbf{s}}_{k}$ generated i.i.d. from Gaussian distribution $\mathcal{CN}(\mathbf{0},\mathbf{Q}_{k})$,~\eqref{58} can be calculated as (see Appendix C)
\begin{equation}\label{59}
   E_{o}(\rho,\mathcal{A}_{l}) = -n\log\mathbb{E}_{\mathbf{H}}\Big\{
  \det\Big(\mathbf{I}_{N_{R}}+\frac{1}{1+\rho}\mathbf{G}_{\mathcal{A}_{l}}\Big)^{-\rho}\Big\},
\end{equation}
where the covariance matrix $\mathbf{G}_{\mathcal{A}_{l}}=\sum \limits_{k \in \mathcal{A}_{l}}
  \mathbf{H}_{k}\mathbf{Q}_{k}\mathbf{H}^{\dagger}_{k}$.

\emph{Proposition 2}: Denote the per channel use error exponent as
\begin{equation}\label{60}
E_{r}(\rho, \mathcal{A}_{l}) = \frac{1}{n}\Bigg[E_{o}(\rho,\mathcal{A}_{l})-\rho\sum_{k \in \mathcal{A}_{l}}R_{k}(n) -\log\binom{k_{n}}{\tilde{k}}\Bigg].
\end{equation}
For $\epsilon \in [0,1]$, if the message-length rate $R_{k}(n)$ is given by
\begin{equation}\label{61}
  R_{k}(n) = (1-\epsilon)c_{k}\mathbb{E}_{\mathbf{H}}\Big\{\log\det\Big(\mathbf{I}_{N_{R}}+\sum \limits_{t \in \mathcal{A}}
  \mathbf{H}_{t}\mathbf{Q}_{t}\mathbf{H}^{\dagger}_{t}\Big)\Big\},
\end{equation}
for the constants $c_{k}>0$ such that $\sum_{k \in \mathcal{A}}c_{k}=n$. Then there exists a positive constant $c_{0}>0$ such that,
\begin{equation}\label{62}
  E_{r}(\rho, \mathcal{A}_{l}) \geq c_{0},
\end{equation}
holds for all sufficient large codelength $n$.

\emph{Proof}: see Appendix D.

Given \emph{Proposition 2}, we have $\mathrm{Pr}\big\{\check{\Xi}_{\tilde{k}}\big\}\rightarrow 0$ for $\forall \tilde{k} \in [1, k_{n}]$ as $n\rightarrow\infty$, provided the individual rate $R_{k}$ follows~\eqref{61}.

\subsection{Proof of the Converse}
The proof of the converse to show that for any sequence of codes $(M_{1},M_{2},\ldots, M_{k_{n}},n)$ with $P^{(n)}_{e} \rightarrow 0$ must have $R_{k}(n)=\log M_{k} \leq c_{k}\mathbb{E}_{\mathbf{H}}\Big\{\log\det\big(\mathbf{I}_{N_{R}}+\sum_{t \in \mathcal{A}}\mathbf{H}_{t}\mathbf{Q}_{t}\mathbf{H}^{\dagger}_{t}\big)\Big\}$ for some positive constant $c_{k}$.

Let $\mathbf{W}=\big\{\mathcal{W}_{1},\mathcal{W}_{2},\ldots, \mathcal{W}_{k_{n}}\big\}$ denote the message of $k_{n}$ users. The entropy of messages are computed by
\begin{align}
  \label{63a} H(\mathbf{W})& = H(\mathbf{W}|\check{\mathbf{Y}}^{n},\mathbf{H}^{n})+I(\mathbf{W};\check{\mathbf{Y}}^{n},\mathbf{H}^{n})  \\
  \label{63b} & \leq H(\mathbf{W}|\check{\mathbf{Y}}^{n},\mathbf{H}^{n})+I(\check{\mathbf{S}}^{n};\check{\mathbf{Y}}^{n},\mathbf{H}^{n}),
\end{align}
where $\check{\mathbf{S}}^{n}= \big\{\check{\mathbf{S}}^{n}_{1},\check{\mathbf{S}}^{n}_{2},\ldots, \check{\mathbf{S}}^{n}_{k_{n}}\big\}$, and~\eqref{63b} follows from the data processing inequality since $\mathbf{W}\rightarrow \check{\mathbf{S}}^{n} \rightarrow (\check{\mathbf{Y}}^{n},\mathbf{H}^{n})$ forms a Markov chain, and the mutual information is given by
\begin{equation}\label{64}
  I(\check{\mathbf{S}}^{n};\check{\mathbf{Y}}^{n},\mathbf{H}^{n})\leq n\mathbb{E}_{\mathbf{H}} \Big\{\log\det\Big(\mathbf{I}_{N_{R}}+\sum \limits_{t \in \mathcal{A}}\mathbf{H}_{t}\mathbf{Q}_{t}\mathbf{H}_{t}^{\dagger}\Big)\Big\}.
\end{equation}

By Fano's inequality, the conditional entropy $H(\mathbf{W}|\check{\mathbf{Y}}^{n},\mathbf{H}^{n})$ is upper bounded by
\begin{equation}\label{65}
  H(\mathbf{W}|\check{\mathbf{Y}}^{n},\mathbf{H}^{n}) \leq 1+P^{(n)}_{e}\sum\limits_{k \in \mathcal{A}}\log M_{k}.
\end{equation}

For the uniformly distributed messages, the entropy is given by $H(\mathbf{W}) = \sum_{k \in \mathcal{A}}\log M_{k} $. Combining~\eqref{63b},~\eqref{64}, and~\eqref{65}, we obtain
\begin{align}\label{66}
   \sum\limits_{k \in \mathcal{A}}\log M_{k}\leq
  n\mathbb{E}_{\mathbf{H}}\Big\{\log\det\Big(\mathbf{I}_{N_{R}}+\sum \limits_{t \in \mathcal{A}}\mathbf{H}_{t}\mathbf{Q}_{t}\mathbf{H}_{t}^{\dagger}\Big)\Big\}
             + 1 + P^{(n)}_{e}  \sum\limits_{k \in \mathcal{A}} \log  M_{k}.
\end{align}

Multiplying both sides with factor $\mu_{k}^{n}$,
\begin{align}\label{67}
   (1-P^{(n)}_{e})\log M_{k} \leq \mu_{k}^{n} +
  n\mu_{k}^{n} \mathbb{E}_{\mathbf{H}}\Big\{\log\det\Big(\mathbf{I}_{N_{R}}+\sum \limits_{t \in \mathcal{A}}\mathbf{H}_{t}\mathbf{Q}_{t}\mathbf{H}_{t}^{\dagger}\Big)\Big\}
\end{align}
Now let $n \rightarrow \infty$, the term $\mu_{k}^{n}$ vanishes as shown in Section II-C. As a consequence, the rate $R_{k}(n)$ is shown to be
\begin{equation}\label{68}
 \log M_{k} \leq c_{k}\mathbb{E}_{\mathbf{H}}\Big\{\log\det\Big(\mathbf{I}_{N_{R}}+\sum \limits_{t \in \mathcal{A}}\mathbf{H}_{t}\mathbf{Q}_{t}\mathbf{H}_{t}^{\dagger}\Big)\Big\},
\end{equation}
for some positive constant $c_{k}$ which is defined in \emph{Theorem 2}.

\section{Proof of Theorem 3 and the Discussions on Main Results}
\subsection{Proof of Theorem 3}
We show that when $\ell\gg k_{\ell}$, the maximum in~\eqref{11} is achieved at $i=k_{\ell}$. Given $\ell\gg k_{\ell}$ and $i=k_{\ell}$,~\eqref{11} behaves as
\begin{equation}\label{69}
   O\big(\ell H_{2}\big(k_{\ell}/\ell\big)/\log k_{\ell}\big).
\end{equation}
On the other hand, when $1\leq i<k_{\ell}$, we have
\begin{equation}\label{70}
  \frac{\log\binom{\ell-k_{\ell}}{i}}{I(\mathbf{y};\mathbf{s}_{\mathcal{A}_{\mathrm{md}}}|\mathbf{s}_{\mathcal{A}_{\mathrm{eq}}},\mathbf{H})}
  =O\big(\ell H_{2}\big(i/\ell\big)/\log i\big).
\end{equation}

We now show the ratio between~\eqref{70} and~\eqref{69} is always less than 1, i.e.,
\begin{equation}\label{71}
  \frac{H_{2}\big(i/\ell\big)}{H_{2}\big(k_{\ell}/\ell\big)}\cdot\frac{\log k_{\ell}}{\log i} < 1, \, \forall i \in[1,k_{\ell}).
\end{equation}

Since $\ell\gg k_{\ell}>i$, we have
\begin{align}
  \label{71a}H_{2}(x)'&=\big(-x\log x-(1-x)\log(1-x)\big)'\\
  \label{71b}&=\log\big((1-x)/x\big),
\end{align}
for $0<x\ll 1/2$. In contrast, the first order derivative of $\log(y)$ is given by $(\log y)'=1/y$,
for $1\leq y \leq k_{\ell}$. In other words, the binary entropy $H_{2}(\cdot)$ increases at a much faster speed than $\log$ function when $i$ goes to $k_{\ell}$, thus the L.H.S. of~\eqref{71} is proved to be lower than $1$ for any $i < k_{\ell}$. As a consequence, the maximum of~\eqref{11} is shown to be achieved at $i=k_{\ell}$. A similar analysis is also applicable for the converse part of \emph{Theorem 1}. By~\eqref{11} and~\eqref{12}, and noting that $\ell\gg k_{\ell}$, we have

\begin{equation}\label{72c}
  n_{0} = \frac{\log\binom{\ell}{k_{\ell}}}{\mathbb{E}_{\mathbf{H}}\Big\{\log\det\big(\mathbf{I}_{N_{R}}+\sum_{k \in \mathcal{A}}\mathbf{H}_{k}\mathbf{Q}_{k}\mathbf{H}^{\dagger}_{k}\big)\Big\}}.
\end{equation}

To adapt to the notation in \emph{Theorem 3}, we replace $\ell$ and $k_{\ell}$ with $\ell_{n}$ and $k_{n}$, respectively. By the inequality $\log\binom{\ell_{n}}{k_{n}}\leq \ell_{n} H_{2}(\frac{k_{n}}{\ell_{n}})$, and combining \emph{Theorem 1} and \emph{2}, the message-length capacity for UE $k$ is lower bounded as
\begin{align}\label{72}
  \nonumber&\lim_{n\rightarrow\infty}\mathbb{E}_{\mathbf{H}}\Big\{\mu_{k}^{(n)}(n-n_{0})\log\det\big(\mathbf{I}_{N_{R}}+\sum_{k \in \mathcal{A}}\mathbf{H}_{k}\mathbf{Q}_{k}\mathbf{H}^{\dagger}_{k}\big)\Big\}\\
  =&-\mu_{k}^{(n)}\log\binom{\ell_{n}}{k_{n}}+c_{k}\mathbb{E}_{\mathbf{H}}\Big\{\log\det\big(\mathbf{I}_{N_{R}}+\sum_{k \in \mathcal{A}}\mathbf{H}_{k}\mathbf{Q}_{k}\mathbf{H}^{\dagger}_{k}\big)\Big\}\\
  \geq & -\mu_{k}^{(n)}\ell_{n} H_{2}\Big(\frac{k_{n}}{\ell_{n}}\Big)
  +c_{k}\mathbb{E}_{\mathbf{H}}\Big\{\log\det\big(\mathbf{I}_{N_{R}}+\sum_{k \in \mathcal{A}}\mathbf{H}_{k}\mathbf{Q}_{k}\mathbf{H}^{\dagger}_{k}\big)\Big\}\\
  = &\, B_{k}(n),
\end{align}
Given the above inequalities, the asymmetric capacity $B_{k}(n)$ in~\eqref{15} is therefore asymptotically achievable.

\subsection{Discussions}
\begin{figure}
  \centering
  \includegraphics[scale=0.25]{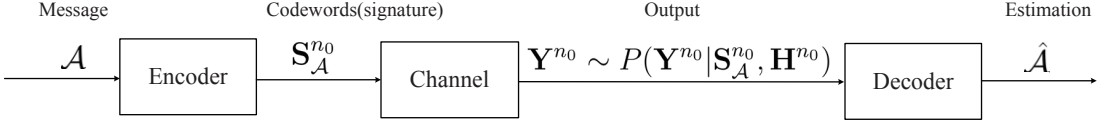}
  \caption{Interpretation of compressed sensing from the view of communication}\label{Fig.3}
\end{figure}
The \emph{Corollary 1} in Section II follows a similar way as in \emph{Proposition 2} by using Kolmogorov's strong law of large numbers~\cite{Sen93}. Hence, all the theorems and corollary in Section II have been established. We now make some comments to those results as follows.
\begin{enumerate}
  \item \emph{Theorem 1} characterizes the asymptotic cost for user identification in MIMO massive access channel. In Fig.~\ref{Fig.3}, we interpret the process of user identification from the view of communication, where the unknown user set $\mathcal{A}$ can be regarded as the information source. The encoder codes the messages $\mathcal{A}$ by using signature symbols and hence obtain the codewords $\mathbf{S}_{\mathcal{A}}$. When receiving the signals from channels, the decoder outputs an estimation to set $\mathcal{A}$ based on received signals $\mathbf{Y}$ and signature symbols $\mathbf{S}$. The problem left is to find the communication times between the transceivers, i.e., $n_{0}$ to ensure the decoder can output an error-free result asymptotically. Based on the set relationship in Fig.~\ref{Fig.2}, Eq.~\eqref{11} in \emph{Theorem 1} follows by taking consideration that when decoder has decoded the subset $\mathcal{A}_{\mathrm{eq}}$ correctly, the number of remain uncertainty sets becomes $|\mathcal{A}_{\mathrm{fa}}|=\binom{\ell-k_{\ell}}{i}$, which means that the decoder's remaining uncertainty about set $\mathcal{A}$ now equals $\log\binom{\ell-k_{\ell}}{i}$. Those uncertainties can be eliminated by the communications between the transceivers through signature symbols. Thus, given the communication rate equals the conditional mutual information
      $I(\mathbf{y};\pmb{\phi}_{\mathcal{A}_{\mathrm{md}}}|\pmb{\phi}_{\mathcal{A}_{\mathrm{eq}}})=\log\det\big(\mathbf{I}_{N_{R}}+\sum\nolimits_{k \in \mathcal{A}_{\mathrm{md}}}\mathbf{H}_{k}\mathbf{Q}_{k}\mathbf{H}^{\dagger}_{k}\big)$, the number of communication times, i.e., the signature length $n_{0}$ is formulated as the ratio between two quantities in~\eqref{11} when set $\mathcal{A}_{\mathrm{eq}}$ is known by decoder. To ensure that all $\mathcal{A}_{\mathrm{eq}}\subseteq \mathcal{A}$ are applicable, a maximum operation is further conducted. The interpretation for the result in~\eqref{12} is akin to that in~\eqref{11}. A detailed discussion on the relationship between multiple access channel and compressed sensing can be also found in~\cite{Jin11}.
  \item \emph{Theorem 2} shows the asymptotic achievable rate in MIMO massive access channels when the active user set $\mathcal{A}$ is known perfectly by receiver. The individual rate in~\eqref{13} is formulated as the sum rate multiplied by a factor $c_{k}=\lim_{n\rightarrow\infty} n\mu_{k}$, where $\mu_{k}$ depends on the codebook size $M_{k}$ for user $k$, and corresponds to the rate allocation for user $k$. Therefore, unlike conventional MAC, the achievable rates in MIMO massive access channels are determined by the codebook sizes for every user instead of SINR. The rate in conventional MAC is achieved by successive decoding at receiver. This approach, however, may not be applicable given finite number of receiving antennas since the interferences now grows unbounded with code codelength. The condition that successive decoding works will be discussed in Section V-C. Combining \emph{Theorem 1} and \emph{Theorem 2}, the achievable message-length capacity region for MIMO massive random access channels is formulated in \emph{Theorem 4}. The finite demission region is rewritten here for the ease of discussion,
      \begin{align}\label{73}
        \nonumber \mathcal{C}_{\mathcal{MAC}}  \triangleq  \Bigg\{&\big(V_{1}(n),...,V_{J}(n)\big):
        \sum_{j=1}^{J}K_{j}V_{j}(n) \\& \leq n\mathbb{E}_{\mathbf{H}}\Big\{\log\det\big(\mathbf{I}_{N_{R}}+\sum_{k \in \mathcal{A}}\mathbf{H}_{k}\mathbf{Q}_{k}\mathbf{H}^{\dagger}_{k}\big)\Big\}
        -\ell_{n} H_{2}(\alpha_{n})\Bigg\}.
      \end{align}
      In contrast to conventional MAC region, the finite demission region~\eqref{73} contains sum constraint only instead of $2^{k_{n}-1}$ rate constraints which corresponds successive decoding. Further, the result in~\eqref{73} indicates that the sum rate of massive random access channel consists of two parts, where the first part denotes the achievable rate when the user activity set $\mathcal{A}$ is known perfectly by the receiver, while the entropy in second part corresponds to the total uncertainty about $\ell_{n}$ users' activity. As a penalty to the unknown user set $\mathcal{A}$, those entropies should be subtracted from the sum capacity.
    \begin{figure}
     \subfigure[$\ell_{n}=n$]{
     \label{Fig.4a} 
     \includegraphics[scale=0.4]{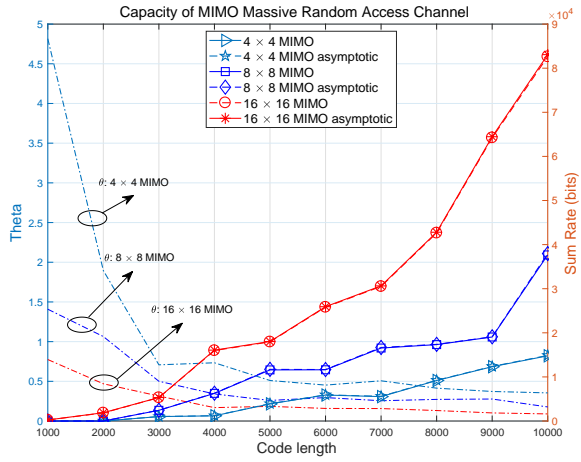}}
     \hspace{0.05in}
     \subfigure[$\ell_{n}=n^{2}$]{
     \label{Fig.4b} 
     \includegraphics[scale=0.4]{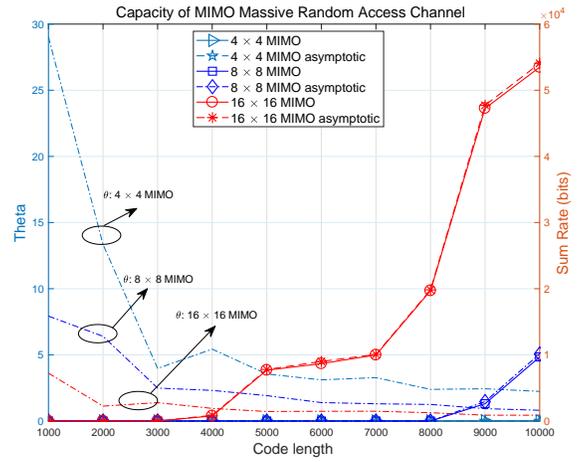}}
     \caption{The capacity of MIMO massive random access channel with coding blocklength}
     \label{Fig.4} 
   \end{figure}

   \begin{figure}
     \subfigure[$\ell_{n}=n$]{
     \label{Fig.5a} 
     \includegraphics[scale=0.4]{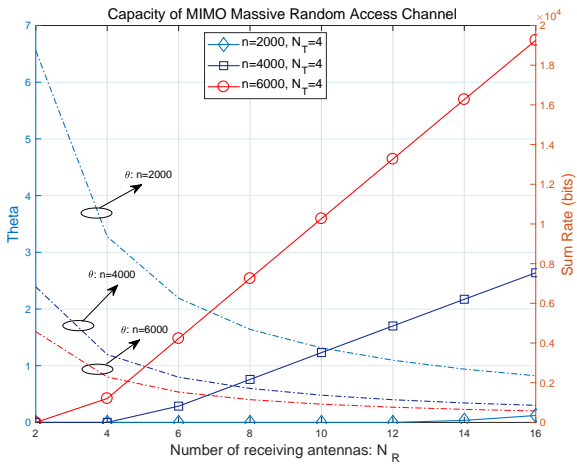}}
     \hspace{0.05in}
     \subfigure[$\ell_{n}=n^{2}$]{
     \label{Fig.5b} 
     \includegraphics[scale=0.4]{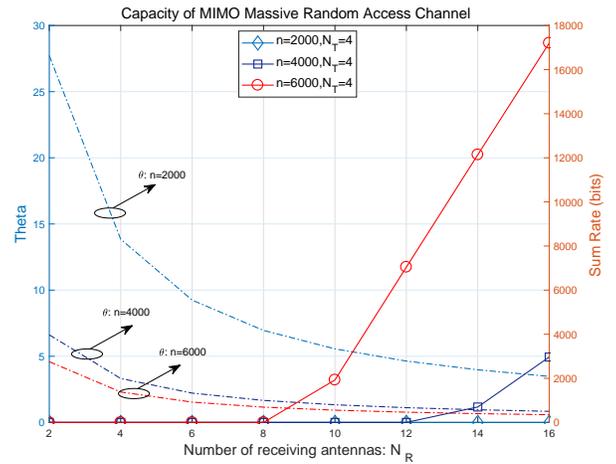}}
     \caption{The capacity of MIMO massive random access channel with the number of receiving antennas}
     \label{Fig.5} 
   \end{figure}
  \item\emph{Corollary 1} indicates that as the codelength increases, the individual rate in MIMO massive access channel will converge to the result that only statistic knowledge of channels is available at transmitter due to the channel hardening effect. In Fig.~\ref{Fig.4}, we plot the sum capacity versus codelength for MIMO massive random access channel. The figures have verified that the asymptotic rate in~\eqref{14}, which depends on statistic knowledge of channels only matches well with the operation rate in~\eqref{13}. The figures also indicate that the sum rates grow unbounded with codelength as well. This is due to fact that we have adopted message-length capacity and assumed that the number of users may vary with codelength. In addition, we have also demonstrated the impact of user identification cost on the capacity of MIMO massive random access in Fig.~\ref{Fig.4} by plotting the dashed curves for $\theta$, which is the ratio between signature length $n_{0}$ and the codelength $n$, and is calculated by Eq.~\eqref{14a}. As figure~\ref{Fig.4} shows, when the coding blocklength $n$ is short, the number of signature required to identify all the active users within the network may exceed the former, which lead to $\theta \geq 1$ and hence the data transmission becomes impossible now. The situation is even severe when $\ell_{n}=n^{2}$ in Fig.~\ref{Fig.4b} since the total uncertainty about user activity $\ell_{n}H_{2}(\alpha_{n})$ is much larger than that when $\ell_{n}=n$. On the other hand, as the codelength increases, the data transmission becomes possible when $\theta < 1$, i.e., there exist remaining channel uses available for data transmission. Increasing the number of receiving antennas $N_{R}$ will also help to identify the active users and raise the channel capacity as shown in Fig.~\ref{Fig.5}. For MIMO massive random access channel given by~\eqref{02} and~\eqref{05}, the degree of freedom (DoF) gain~\cite{Shin03} is shown to be $N_{\mathrm{DoF}}=\min\{N_{R},k_{n}N_{T}\}=N_{R}$, which indicates that the individual rate will increase linearly with the number of receiving antennas. The results are verified in Fig.~\ref{Fig.5}, where as the $N_{R}$ increases, $\theta$ becomes lower down and when it drops below $1$, the capacities tend to increase linearly with $N_{R}$ for both $\ell_{n}=n$ and $\ell_{n}=n^{2}$.
\end{enumerate}

\subsection{Successive Decoding}
In previous discussion, we mentioned that the rate for conventional MAC is not achievable for MIMO massive access channel since successive decoding may not be applicable. On the other hand, since multiple antennas are deployed at the receiver, a natural question is that can we use multiple antenna to combat the large user interferences? We begin the analysis of successive decoding with multiple antennas in this section, and assume the number of users is $k_{n}=O(n)$ with no random user activity. We consider the successive interference cancellation (SIC) where user $k_{n}$ is decoded first. The capacity of SIC for user $k_{n}$ is given by
\begin{align}
   \label{74a} C_{k_{n}}&= I(\check{\mathbf{Y}},\mathbf{H};\check{\mathbf{S}})
   -I(\check{\mathbf{Y}},\mathbf{H};\check{\mathbf{S}}_{x_{1}},\ldots,\check{\mathbf{S}}_{x_{k_{n}-1}}|\check{\mathbf{S}}_{x_{k_{n}}})\\
   \nonumber &=\mathbb{E}_{\mathbf{H}}\Bigg\{\log\det\Big(\mathbf{I}_{N_{R}}+\sum_{k \in \mathcal{A}}\mathbf{H}_{k}\mathbf{Q}_{k}\mathbf{H}^{\dagger}_{k}\Big)\\
   \label{74b}&\qquad \qquad -\log\det\Big(\mathbf{I}_{N_{R}}+\sum_{k \in \mathcal{A}\backslash\{k_{n}\}}\mathbf{H}_{k}\mathbf{Q}_{k}\mathbf{H}^{\dagger}_{k}\Big)\Bigg\}\\
   \label{74c}&\xrightarrow{k_{n}\rightarrow\infty} N_{R}\log\Big(1+\frac{\beta_{k_{n}}p_{k_{n}}}{1+\sum_{k\in \mathcal{A}\backslash \{k_{n}\}}\beta_{k}p_{k}}\Big),
\end{align}
where $\check{\mathbf{S}}=\{\check{\mathbf{S}}_{x_{1}},\ldots,\check{\mathbf{S}}_{x_{k_{n}}}\}$.

In the seminar work~\cite{Chen17}, it was shown that the error probability of successive interference cancellation for the first user decays at the rate of $\exp\{-\delta n C_{k_{n}}\}$, where $\delta$ is some positive constant. When the number of received antenna $N_{R}$ is finite, $n C_{k_{n}}$ converges to some constant such that the error probability is not guaranteed to vanish, i.e.,
\begin{align}
    \label{82d} \lim_{n\rightarrow\infty} nC_{k_{n}}&= \lim_{n\rightarrow\infty}
    nN_{R}\log\Big(1+\frac{\beta_{k_{n}}p_{k_{n}}}{1+\sum_{k\in \mathcal{A}\backslash \{k_{n}\}}\beta_{k}p_{k}}\Big)\\
    \label{75a}&\approx \lim_{n\rightarrow\infty} N_{R}\frac{n\beta_{k_{n}}p_{k_{n}}}{1+\sum_{k\in \mathcal{A}\backslash \{k_{n}\}}\beta_{k}p_{k}}
\end{align}
converges as $n$ grows, where~\eqref{75a} follows from $\log(1+ x) \approx x$ for $x\rightarrow 0$, Thus, the lower bound of successive decoding error probability is not guaranteed to be vanished when the number of receiving antenna $N_{R}$ is finite, and the successive decoding is never applicable for MIMO massive access channel.

However, the situation becomes different if the number of received antenna $N_{R}$ is comparable with the codelength $n$. In what follows, we derive the upper bound of error probability by successive decoding based on dependence-testing (DT) bound~\cite{Polyanskiy10}. Let $k_{n}$ to be the first decoded user. The code rates of user $k_{n}$ is given by
\begin{align}\label{76}
   R_{k_{n}}&=\frac{1}{n}\log M_{k_{n}}\\
  &=(1-\epsilon)\mathbb{E}_{\mathbf{H}}\Big\{\log\det\Big[\mathbf{I}_{N_{R}}+\mathbf{G}_{\bar{k}_{n}}^{-1}\big(\mathbf{H}_{k_{n}}\mathbf{Q}_{k_{n}}\mathbf{H}^{\dagger}_{k_{n}}\big)
  \Big]\Big\},
\end{align}
where $\mathbf{G}_{\bar{k}_{n}}=\mathbf{I}_{N_{R}}+\sum_{k \in \mathcal{A}\setminus\{k_{n}\}}
  \mathbf{H}_{k}\mathbf{Q}_{k}\mathbf{H}^{\dagger}_{k}$.
By DT bound, the error probability of successive decoding is upper bounded as
\begin{align}\label{77}
  P_{e}^{(n)}\leq \mathbb{E}\Big[\exp\Big\{-\Big(\imath(\check{\mathbf{Y}},\mathbf{H};\check{\mathbf{S}}_{k_{n}})-\log\frac{ M_{kn}-1}{2}\Big)^{+}\Big\}\Big],
\end{align}
where $x^{+}=\max(x,0)$, and the information density is calculated as
\begin{align}\label{78}
  \nonumber\imath(\check{\mathbf{Y}},&\mathbf{H};\check{\mathbf{S}}_{k_{n}})=
  -\sum_{i=1}^{n}\Big[\check{\mathbf{y}}_{i}^{\dagger}\Big(\mathbf{I}_{N_{R}}+\sum_{k \in \mathcal{A}}\mathbf{H}_{k}\mathbf{Q}_{k}\mathbf{H}^{\dagger}_{k}\Big)^{-1}\check{\mathbf{y}}_{i}
  -(\check{\mathbf{y}}_{i}-\mathbf{H}_{k_{n}}\check{\mathbf{s}}_{k_{n},i})^{\dagger}\mathbf{G}_{\bar{k}_{n}}^{-1}
 (\check{\mathbf{y}}_{i}-\mathbf{H}_{k_{n}}\check{\mathbf{s}}_{k_{n},i})\Big]\\
 &\qquad \qquad \qquad \qquad
 +n\log\det\Big[\mathbf{I}_{N_{R}}+\mathbf{G}_{\bar{k}_{n}}^{-1}\big(\mathbf{H}_{k_{n}}\mathbf{Q}_{k_{n}}\mathbf{H}^{\dagger}_{k_{n}}\big)\Big],
\end{align}
where $\check{\mathbf{y}}_{i}$ and $\check{\mathbf{s}}_{k_{n},i}$ denote the received signal and the codewords of user $k_{n}$ in $i^{th}$ channel use, respectively. By the law of large numbers and the i.i.d. of signals in each channel use, and noting that
\begin{equation}\label{79}
\mathbb{E}\{\check{\mathbf{y}}_{i}\check{\mathbf{y}}_{i}^{\dagger}\}=\mathbf{I}_{N_{R}}+\sum_{k \in \mathcal{A}}\mathbf{H}_{k}\mathbf{Q}_{k}\mathbf{H}^{\dagger}_{k},
\end{equation}
and
\begin{align}\label{80}
\mathbb{E}\big\{(\check{\mathbf{y}}_{i}-\mathbf{H}_{k_{n}}\check{\mathbf{s}}_{k_{n},i})(\check{\mathbf{y}}_{i}-\mathbf{H}_{k_{n}}\check{\mathbf{s}}_{k_{n},i})^{\dagger}\big\}
=\mathbf{I}_{N_{R}}+\sum_{k \in \mathcal{A}\setminus\{k_{n}\}}\mathbf{H}_{k}\mathbf{Q}_{k}\mathbf{H}^{\dagger}_{k},
\end{align}
we have
\begin{align}\label{81}
  \nonumber &\lim_{n\rightarrow\infty} \sum_{i=1}^{n}\check{\mathbf{y}}_{i}^{\dagger}\Big(\mathbf{I}_{N_{R}}+\sum_{k \in \mathcal{A}}\mathbf{H}_{k}\mathbf{Q}_{k}\mathbf{H}^{\dagger}_{k}\Big)^{-1}\check{\mathbf{y}}_{i}\\
=&\,\lim_{n\rightarrow\infty}\mathrm{Tr}\Bigg\{\Big(\mathbf{I}_{N_{R}}+\sum_{k \in \mathcal{A}}\mathbf{H}_{k}\mathbf{Q}_{k}\mathbf{H}^{\dagger}_{k}\Big)^{-1}\sum_{i=1}^{n}\check{\mathbf{y}}_{i}\check{\mathbf{y}}_{i}^{\dagger}\Bigg\}\\
=&\, nN_{R}.
\end{align}
In a similar approach,
\begin{align}\label{82}
  \lim_{n\rightarrow\infty} \sum_{i=1}^{n}
(\check{\mathbf{y}}_{i}-\mathbf{H}_{k_{n}}&\check{\mathbf{s}}_{k_{n},i})^{\dagger}\mathbf{G}_{\bar{k}_{n}}^{-1}(\check{\mathbf{y}}_{i}-\mathbf{H}_{k_{n}}\check{\mathbf{s}}_{k_{n},i}) = nN_{R}.
\end{align}

Thus, the exponent is given by
\begin{align}
  \nonumber&\lim_{n\rightarrow\infty}\imath(\check{\mathbf{Y}},\mathbf{H};\check{\mathbf{S}}_{k_{n}})-\log\frac{ M_{kn}-1}{2}\\
\label{83a}=\,&\epsilon N_{R}n\log\Big(1+\frac{\beta_{k_{n}}p_{k_{n}}}{1+\sum_{k\in \mathcal{A}\backslash \{k_{n}\}}\beta_{k}p_{k}}\Big)+\log2\\
\label{83b}\approx\, & \epsilon N_{R}\frac{n\beta_{k_{n}}p_{k_{n}}}{1+\sum_{k\in \mathcal{A}\backslash \{k_{n}\}}\beta_{k}p_{k}}+\log2,
\end{align}
where~\eqref{83b} follows from $\log(1+ x) \approx x$ for small $x$. Since $k_{n}=O(n)$ and $p_{k}=\Theta(1)$, it can be observed that the exponent converges as codelength $n$ grows when $N_{R}$ is some constant. On the other hand, when $N_{R}=O(n)$ either, the exponent grows unbounded with $n$ and decode error probability can thus be made arbitrary small.

\section{Conclusion}
In this paper, the achievable capacity region of MIMO massive access channel is investigated. In contrast to conventional MAC model, the number of simultaneous communicating users may grow unbounded with codelength $n$, thus lead a different behaviour for the fundamental limits. Specifically, we showed that the asymptotic rate for each user can be formulated as the sum capacity multiplied by a specific factor corresponds to the rate allocation. Thus, the rate for conventional MAC due to successive decoding is never applied. Further, the random users access is also considered, and the asymptotic users identification cost is quantified by using concentration inequalities which are related to the information densities of transmit signatures. The analysis also suggests that successive decoding is possible when the number of receiving antenna is also comparable with coding blocklength $n$.

%

\appendices
\section{Proof of Lemma 1}
The conditional information density is given by
\begin{equation}\label{A1}
  \imath(\mathbf{Y};\mathbf{S}_{\mathcal{A}_{\mathrm{md}}}|\mathbf{S}_{\mathcal{A}_{\mathrm{eq}}},\mathbf{H}) =
  \log\frac{P(\mathbf{Y}|\mathbf{S}_{\mathcal{A}_{\mathrm{md}}},\mathbf{S}_{\mathcal{A}_{\mathrm{eq}}},\mathbf{H})}
  {P(\mathbf{Y}|\mathbf{S}_{\mathcal{A}_{\mathrm{eq}}},\mathbf{H})}.
\end{equation}
Due to the i.i.d. generation of the signature symbols and the memoryless nature of channel,~\eqref{A1} can be rewritten as
\begin{align}
  \label{A2a}\imath(\mathbf{Y};\mathbf{S}_{\mathcal{A}_{\mathrm{md}}}|\mathbf{S}_{\mathcal{A}_{\mathrm{eq}}},\mathbf{H}) &=
  \log\frac{P(\mathbf{Y}|\mathbf{S}_{\mathcal{A}_{\mathrm{md}}},\mathbf{S}_{\mathcal{A}_{\mathrm{eq}}},\mathbf{H})}
  {P(\mathbf{Y}|\mathbf{S}_{\mathcal{A}_{\mathrm{eq}}},\mathbf{H})}\\
  \label{A2b}&=\sum_{i=1}^{n_{0}}\log\frac{P(\mathbf{y}(i)|\mathbf{s}_{\mathcal{A}_{\mathrm{md}}}(i),\mathbf{s}_{\mathcal{A}_{\mathrm{eq}}}(i),\mathbf{H}(i))}
  {P(\mathbf{y}(i)|\mathbf{s}_{\mathcal{A}_{\mathrm{eq}}}(i),\mathbf{H}(i))},
\end{align}
where the probabilities are given by
\begin{equation}\label{A3}
  P(\mathbf{y}(i)|\mathbf{s}_{\mathcal{A}_{\mathrm{md}}}(i),\mathbf{s}_{\mathcal{A}_{\mathrm{eq}}}(i),\mathbf{H}(i))
  =\mathcal{CN}(\mathbf{z}(i);\mathbf{0},\mathbf{I}_{N_{R}}),
\end{equation}
and
\begin{equation}\label{A4}
  P(\mathbf{y}(i)|\mathbf{s}_{\mathcal{A}_{\mathrm{eq}}}(i),\mathbf{H}(i))=
  \mathcal{CN}\Big(\mathbf{z}(i)+\sum_{k\in\mathcal{A}_{\mathrm{md}}}\mathbf{H}_{k}(i)\mathbf{s}_{k}(i);\mathbf{0},\mathbf{I}_{N_{R}}
  +\mathbf{G}_{\mathcal{A}_{\mathrm{md}}}(i)\Big),
\end{equation}
where
\begin{equation}\label{A5}
  \mathbf{G}_{\mathcal{A}_{\mathrm{md}}}(i)=\sum_{k \in \mathcal{A}_{\mathrm{md}}}\mathbf{H}_{k}(i)\mathbf{Q}_{k}\mathbf{H}^{\dagger}_{k}(i).
\end{equation}
Combining~\eqref{A2b},~\eqref{A3}, and~\eqref{A4}, we obtain
\begin{align}\label{A6}
  \nonumber\imath(\mathbf{y}(i);\mathbf{s}_{\mathcal{A}_{\mathrm{md}}}(i)|\mathbf{s}_{\mathcal{A}_{\mathrm{eq}}}(i),\mathbf{H}(i))
  =& \,\hat{i}(\mathbf{y}(i);\mathbf{s}_{\mathcal{A}_{\mathrm{md}}}(i)|\mathbf{s}_{\mathcal{A}_{\mathrm{eq}}}(i),\mathbf{H}(i))
  -\|\mathbf{z}(i)\|^{2}\\
  +(\mathbf{z}(i)+\pmb{\phi}_{\mathcal{A}_{\mathrm{md}}}(i)&)^{\dagger}
  \big(\mathbf{I}_{N_{R}}+\mathbf{G}_{\mathcal{A}_{\mathrm{md}}}(i)\big)^{-1}
  (\mathbf{z}(i)+\pmb{\phi}_{\mathcal{A}_{\mathrm{md}}}(i)),
\end{align}
where $\hat{i}(\mathbf{y}(i);\mathbf{s}_{\mathcal{A}_{\mathrm{md}}}(i)|\mathbf{s}_{\mathcal{A}_{\mathrm{eq}}}(i),\mathbf{H}(i))
=\log\det(\mathbf{I}_{N_{R}}+\mathbf{G}_{\mathcal{A}_{\mathrm{md}}}(i))$, and $\pmb{\phi}_{\mathcal{A}_{\mathrm{md}}}(i)=\sum\limits_{k\in\mathcal{A}_{\mathrm{md}}}\mathbf{H}_{k}(i)\mathbf{s}_{k}(i)$.

For notational simplification, we will drop the time index $i$ during the subsequent derivations. Define $\mathbf{\Omega} _{\mathcal{A}_{\mathrm{md}}}\triangleq\mathbf{I}_{N_{R}}+\mathbf{G}_{\mathcal{A}_{\mathrm{md}}}$, and
\begin{align}\label{A7}
  W &\triangleq \imath(\mathbf{y};\mathbf{s}_{\mathcal{A}_{\mathrm{md}}}|\mathbf{s}_{\mathcal{A}_{\mathrm{eq}}},\mathbf{H})
  -I(\mathbf{y};\mathbf{s}_{\mathcal{A}_{\mathrm{md}}}|\mathbf{s}_{\mathcal{A}_{\mathrm{eq}}},\mathbf{H})\\
  &=-I(\mathbf{y};\mathbf{s}_{\mathcal{A}_{\mathrm{md}}}|\mathbf{s}_{\mathcal{A}_{\mathrm{eq}}},\mathbf{H})
  +\hat{i}(\mathbf{y};\mathbf{s}_{\mathcal{A}_{\mathrm{md}}}|\mathbf{s}_{\mathcal{A}_{\mathrm{eq}}},\mathbf{H})
  -\|\mathbf{z}\|^{2}+(\mathbf{z}+\pmb{\phi}_{\mathcal{A}_{\mathrm{md}}})^{\dagger}
  \mathbf{\Omega}_{\mathcal{A}_{\mathrm{md}}}^{-1}(\mathbf{z}+\pmb{\phi}_{\mathcal{A}_{\mathrm{md}}})\\
  \nonumber &=  -I(\mathbf{y};\mathbf{s}_{\mathcal{A}_{\mathrm{md}}}|\mathbf{s}_{\mathcal{A}_{\mathrm{eq}}},\mathbf{H})
  +\hat{i}(\mathbf{y};\mathbf{s}_{\mathcal{A}_{\mathrm{md}}}|\mathbf{s}_{\mathcal{A}_{\mathrm{eq}}},\mathbf{H})
  -\mathbf{z}^{\dagger}\big(\mathbf{I}_{N_{R}}
  -\mathbf{\Omega}_{\mathcal{A}_{\mathrm{md}}}^{-1}\big)\mathbf{z}\\
  &\quad+\pmb{\phi}_{\mathcal{A}_{\mathrm{md}}}^{\dagger}\mathbf{\Omega}_{\mathcal{A}_{\mathrm{md}}}^{-1}
  \pmb{\phi}_{\mathcal{A}_{\mathrm{md}}}+\mathbf{z}^{\dagger}\mathbf{\Omega}_{\mathcal{A}_{\mathrm{md}}}^{-1}\pmb{\phi}_{\mathcal{A}_{\mathrm{md}}}
  +\pmb{\phi}_{\mathcal{A}_{\mathrm{md}}}^{\dagger}\mathbf{\Omega}_{\mathcal{A}_{\mathrm{md}}}^{-1}\mathbf{z}\\
  \nonumber&= -I(\mathbf{y};\mathbf{s}_{\mathcal{A}_{\mathrm{md}}}|\mathbf{s}_{\mathcal{A}_{\mathrm{eq}}},\mathbf{H})
  +\hat{i}(\mathbf{y};\mathbf{s}_{\mathcal{A}_{\mathrm{md}}}|\mathbf{s}_{\mathcal{A}_{\mathrm{eq}}},\mathbf{H})
  -\mathbf{z}^{\dagger}\mathbf{\Omega}_{\mathcal{A}_{\mathrm{md}}}^{-1}\mathbf{G}_{\mathcal{A}_{\mathrm{md}}}\mathbf{z}\\
  &\quad+\hat{\pmb{\phi}}_{\mathcal{A}_{\mathrm{md}}}^{\dagger}\mathbf{\Omega}_{\mathcal{A}_{\mathrm{md}}}^{-1}\mathbf{G}_{\mathcal{A}_{\mathrm{md}}}\hat{\pmb{\phi}}_{\mathcal{A}_{\mathrm{md}}}
  +\mathbf{z}^{\dagger}\mathbf{\Omega}_{\mathcal{A}_{\mathrm{md}}}^{-1}\mathbf{P}\hat{\pmb{\phi}}_{\mathcal{A}_{\mathrm{md}}}
  +\hat{\pmb{\phi}}_{\mathcal{A}_{\mathrm{md}}}^{\dagger}\mathbf{P}^{\dagger}\mathbf{\Omega}_{\mathcal{A}_{\mathrm{md}}}^{-1}\mathbf{z}\\
  \nonumber &\leq -I(\mathbf{y};\mathbf{s}_{\mathcal{A}_{\mathrm{md}}}|\mathbf{s}_{\mathcal{A}_{\mathrm{eq}}},\mathbf{H})
  +\hat{i}(\mathbf{y};\mathbf{s}_{\mathcal{A}_{\mathrm{md}}}|\mathbf{s}_{\mathcal{A}_{\mathrm{eq}}},\mathbf{H})
  -\mathbf{z}^{\dagger}\mathbf{\Omega}_{\mathcal{A}_{\mathrm{md}}}^{-1}\mathbf{G}_{\mathcal{A}_{\mathrm{md}}}\mathbf{z}\\
  &\quad+2\hat{\pmb{\phi}}_{\mathcal{A}_{\mathrm{md}}}^{\dagger}\mathbf{\Omega}_{\mathcal{A}_{\mathrm{md}}}^{-1}\mathbf{G}_{\mathcal{A}_{\mathrm{md}}}
  \hat{\pmb{\phi}}_{\mathcal{A}_{\mathrm{md}}}+\mathbf{z}^{\dagger}\mathbf{\Omega}_{\mathcal{A}_{\mathrm{md}}}^{-1}\mathbf{z},
\end{align}
where by Cholesky decomposition, we have $\mathbf{G}_{\mathcal{A}_{\mathrm{md}}}=\mathbf{P}\mathbf{P}^{\dagger}$, $\pmb{\phi}_{\mathcal{A}_{\mathrm{md}}}=\mathbf{P}\hat{\pmb{\phi}}_{\mathcal{A}_{\mathrm{md}}}$, and $\hat{\pmb{\phi}}_{\mathcal{A}_{\mathrm{md}}}\sim \mathcal{CN}(\mathbf{0},\mathbf{I}_{N_{R}})$. Let
\begin{equation}\label{A8}
  A=-\mathbf{z}^{\dagger}\mathbf{\Omega}_{\mathcal{A}_{\mathrm{md}}}^{-1}\mathbf{G}_{\mathcal{A}_{\mathrm{md}}}\mathbf{z}+2\hat{\pmb{\phi}}_{\mathcal{A}_{\mathrm{md}}}^{\dagger}\mathbf{\Omega}_{\mathcal{A}_{\mathrm{md}}}^{-1}\mathbf{G}_{\mathcal{A}_{\mathrm{md}}}
  \hat{\pmb{\phi}}_{\mathcal{A}_{\mathrm{md}}}+\mathbf{z}^{\dagger}\mathbf{\Omega}_{\mathcal{A}_{\mathrm{md}}}^{-1}\mathbf{z}
\end{equation}
and
\begin{equation}\label{A9}
  B=-I(\mathbf{y};\mathbf{s}_{\mathcal{A}_{\mathrm{md}}}|\mathbf{s}_{\mathcal{A}_{\mathrm{eq}}},\mathbf{H})
  +\hat{i}(\mathbf{y};\mathbf{s}_{\mathcal{A}_{\mathrm{md}}}|\mathbf{s}_{\mathcal{A}_{\mathrm{eq}}},\mathbf{H}),
\end{equation}
with singular value decomposition (SVD), we have
\begin{align}\label{A10}
  |A| & \leq 3\sum_{i=1}^{N_{R}}\sigma_{i}(\mathbf{A})|\hat{Z}_{\mathrm{max}}|^{2}
  +\sum_{i=1}^{N_{R}}\sigma_{i}(\mathbf{B})|\hat{Z}_{\mathrm{max}}|^{2},
\end{align}
where $\mathbf{A}=(\mathbf{I}_{N_{R}}+\mathbf{G}_{\mathcal{A}_{\mathrm{md}}}^{-1})^{-1}$, $\mathbf{B}=(\mathbf{I}_{N_{R}}+\mathbf{G}_{\mathcal{A}_{\mathrm{md}}})^{-1}$, and $\sigma_{i}(\cdot)$ denotes the $i^{th}$ singular value of matrix. Given the unitary matrices $\mathbf{U}$ and $\mathbf{V}$ for the SVD of $\mathbf{A}$ and $\mathbf{B}$, $\hat{Z}_{\mathrm{max}}$ is the maximum absolution value among the entries in vectors $\mathbf{U}\hat{\pmb{\phi}}_{\mathcal{A}_{\mathrm{md}}}$, $\mathbf{Uz}$, and $\mathbf{Vz}$. Let $\hat{Z}_{\mathrm{max}} = |Z^{(r)}_{\mathrm{max}}+jZ^{(j)}_{\mathrm{max}}|$, where $Z^{(r,j)}_{\mathrm{max}}\sim \mathcal{N}(0,1)$. Eq~\eqref{A10} is further upper bounded as
\begin{align}\label{A11}
  \nonumber |A| &\leq 2\sum_{i=1}^{N_{R}}\Big[3\sigma_{i}(\mathbf{A})
  +\sigma_{i}(\mathbf{B})\Big]\max\big\{|Z^{(r)}_{\mathrm{max}}|^{2},|Z^{(j)}_{\mathrm{max}}|^{2}\big\}\\
  &=2\big[3\mathrm{Tr}(\mathbf{A})+\mathrm{Tr}(\mathbf{B})\big]\max\big\{|Z^{(r)}_{\mathrm{max}}|^{2},|Z^{(j)}_{\mathrm{max}}|^{2}\big\}.
\end{align}
The expectation of $q^{th}$ moment of $|W|$ is thus upper bounded as
\begin{align}
  \label{A12a} \mathbb{E}[|A|^{q}] &\leq 2^{q} \mathbb{E}_{\mathbf{H}}\Big[\big(3\mathrm{Tr}(\mathbf{A})+\mathrm{Tr}(\mathbf{B})\big)^{q}\Big] \mathbb{E}\big[(Z^{(r)}_{\mathrm{max}})^{2q}+(Z^{(i)}_{\mathrm{max}})^{2q}\big] \\
  \label{A12b}&\leq 4^{q}\mathbb{E}_{\mathbf{H}}\Big[\big(3\mathrm{Tr}(\mathbf{A})+\mathrm{Tr}(\mathbf{B})\big)^{q}\Big]
  \frac{2}{\sqrt{\pi}}\Gamma(q+\frac{1}{2})\\
  \label{A12c}&\leq 4^{q}\mathbb{E}_{\mathbf{H}}\Big[\big(3\mathrm{Tr}(\mathbf{A})+\mathrm{Tr}(\mathbf{B})\big)^{q}\Big]2q!\\
  \label{A12d}&\leq (16N_{R})^{q}2q!
\end{align}
where~\eqref{A12b} follows from $\mathbb{E}[\hat{Z}^{2q}]\leq \frac{2^{q}}{\sqrt{\pi}}\Gamma(q+\frac{1}{2})$, for the Gaussian random variable $\hat{Z}\sim N(0,1)$,~\eqref{A12c} follows from $\Gamma(q+\frac{1}{2})\leq\sqrt{\pi}q!$,~and~\eqref{A12d} follows since $\mathbf{G}_{\mathcal{A}_{\mathrm{md}}}$ is positive definite, the eigenvalues of matrix $\mathbf{A}$ and $\mathbf{B}$ are strictly lower than $1$.


On the other hand, the expectation of the $q^{th}$ moment of $B$ is given by
\begin{align}
  \label{A13a}\mathbb{E}_{\mathbf{H}}[|B|^{q}] & = \mathbb{E}_{\mathbf{H}}\Big[\big|\hat{i}(\mathbf{y};\mathbf{s}_{\mathcal{A}_{\mathrm{md}}}|\mathbf{s}_{\mathcal{A}_{\mathrm{eq}}},\mathbf{H})
  -I(\mathbf{y};\mathbf{s}_{\mathcal{A}_{\mathrm{md}}}|\mathbf{s}_{\mathcal{A}_{\mathrm{eq}}},\mathbf{H})\big|^{q}\Big]\\
  \label{A13b}&\leq q!\mathbb{E}_{\mathbf{H}}
  \Big\{\exp\Big[\hat{i}(\mathbf{y};\mathbf{s}_{\mathcal{A}_{\mathrm{md}}}|\mathbf{s}_{\mathcal{A}_{\mathrm{eq}}},\mathbf{H})
  -I(\mathbf{y};\mathbf{s}_{\mathcal{A}_{\mathrm{md}}}|\mathbf{s}_{\mathcal{A}_{\mathrm{eq}}},\mathbf{H})\Big]\Big\}\\
  \label{A13c}&=q!\mathbb{E}_{\mathbf{H}}\Big\{\det\big(\mathbf{I}_{N_{R}}+\mathbf{G}_{\mathcal{A}_{\mathrm{md}}}\big)\Big\}
  e^{-I(\mathbf{y};\mathbf{s}_{\mathcal{A}_{\mathrm{md}}}|\mathbf{s}_{\mathcal{A}_{\mathrm{eq}}},\mathbf{H})},
\end{align}
where by choosing $t=1$,~\eqref{A13b} follows from the moment generating function equality
\begin{equation}\label{A14}
  \mathbb{E}_{X}[e^{tX}] \geq \frac{t^{q}}{q!}\mathbb{E}_{X}[X^{q}].
\end{equation}

By the inequality
\begin{equation}\label{A15}
  \mathbb{E}[|A+B|^{q}]\leq 2^{q-1}\big(\mathbb{E}[|A|^{q}]+\mathbb{E}[|B|^{q}]\big), \, q>1,
\end{equation}
and combing~\eqref{A12d} and~\eqref{A13c}, we obtain
\begin{align}\label{A16}
  \mathbb{E}[|W|^{q}]&\leq q!\Big\{(32N_{R})^{q}+2^{q-1}
  \mathbb{E}_{\mathbf{H}}\Big[\det\big(\mathbf{I}_{N_{R}}+\mathbf{G}_{\mathcal{A}_{\mathrm{md}}}\big)\Big]
  e^{-I(\mathbf{y};\mathbf{s}_{\mathcal{A}_{\mathrm{md}}}|\mathbf{s}_{\mathcal{A}_{\mathrm{eq}}},\mathbf{H})}\Big\}\\
  &\leq q!\Big[32N_{R}+
  \mathbb{E}_{\mathbf{H}}\Big[\det\big(\mathbf{I}_{N_{R}}+\mathbf{G}_{\mathcal{A}_{\mathrm{md}}}\big)\Big]
  e^{-I(\mathbf{y};\mathbf{s}_{\mathcal{A}_{\mathrm{md}}}|\mathbf{s}_{\mathcal{A}_{\mathrm{eq}}},\mathbf{H})}\Big]^{q}
\end{align}
By choosing $q=2$, we have
\begin{equation}\label{A17}
  \mathbb{E}[|W|^{2}]\leq 2\Big[32N_{R}+
  \mathbb{E}_{\mathbf{H}}\Big[\det\big(\mathbf{I}_{N_{R}}+\mathbf{G}_{\mathcal{A}_{\mathrm{md}}}\big)\Big]
  e^{-I(\mathbf{y};\mathbf{s}_{\mathcal{A}_{\mathrm{md}}}|\mathbf{s}_{\mathcal{A}_{\mathrm{eq}}},\mathbf{H})}\Big]^{2}
\end{equation}

Thus, setting $c=32N_{R}+
\mathbb{E}_{\mathbf{H}}\Big[\det\big(\mathbf{I}_{N_{R}}+\mathbf{G}_{\mathcal{A}_{\mathrm{md}}}\big)\Big]
  e^{-I(\mathbf{y};\mathbf{s}_{\mathcal{A}_{\mathrm{md}}}|\mathbf{s}_{\mathcal{A}_{\mathrm{eq}}},\mathbf{H})}$, $\tau=2n_{0}c^{2}$, we have
\begin{equation}\label{A18}
  n_{0}\mathbb{E}[|W|^{2}] \leq \tau,
\end{equation}
and
\begin{equation}\label{A19}
  n_{0}\mathbb{E}[|W|^{q}] \leq \frac{q!}{2}\tau c^{q-2},
\end{equation}
which satisfy the conditions for the Bernstein's inequality~\cite{Boucheron13}. By using Bernstein's inequality, we obtain the desired results in \emph{Lemma 1}.

Before ending the proving of the proposition, we show that $c$ is some positive and finite constant. Since $c$ is given by
\begin{equation}\label{A20}
  c=32N_{R}+\mathbb{E}_{\mathbf{H}}
  \Big[\det\big(\mathbf{I}_{N_{R}}+\mathbf{G}_{\mathcal{A}_{\mathrm{md}}}\big)\Big]
  e^{-I(\mathbf{y};\mathbf{s}_{\mathcal{A}_{\mathrm{md}}}|\mathbf{s}_{\mathcal{A}_{\mathrm{eq}}},\mathbf{H})},
\end{equation}
it follows directly $c$ is some positive constant when the set cardinality $|\mathcal{A}_{\mathrm{md}}|$ is some finite constant. On the other hand, when $|\mathcal{A}_{\mathrm{md}}|$ grow sublinearly or linearly with $n$, we have (see~\eqref{D4}-\eqref{D7})
\begin{equation}\label{A21}
      \lim \limits_{|\mathcal{A}_{\mathrm{md}}|\rightarrow \infty} \det\Big(\mathbf{I}_{N_{R}}+\sum_{k \in \mathcal{A}_{\mathrm{md}}}\mathbf{H}_{k}\mathbf{Q}_{k}\mathbf{H}^{\dagger}_{k}\Big)
      = \Big(1+  \sum\limits_{t \in \mathcal{A}_{\mathrm{md}}}\beta_{t}\mathrm{Tr}(\mathbf{Q}_{t})\Big)^{N_{R}},
\end{equation}
which depends on the statistical knowledge of channel $\mathbf{H}$ only. Therefore, we may conclude that the limits of $c$ equals $32N_{R}+1$ in the latter case.

\section{Proof of Proposition 1}
We now show the condition in~\eqref{47} holds asymptotically if~\eqref{48} is satisfied by considering the separated cases for the cardinalities of set $\mathcal{A}_{\mathrm{md}}$. Note that $c$ is some positive constant as shown in Appendix A.

\emph{Case a)}: $i=|\mathcal{A}_{\mathrm{md}}|$ is some constant. Since $\log\binom{p}{k}=O\big(k\log\frac{p}{k}+(p-k)\log\frac{p}{p-k}\big)$, the L.H.S. of~\eqref{47} is given by
\begin{align}\label{B1}
  \nonumber&\frac{n_{0}\big[\delta_{2}
  I(\mathbf{y};\mathbf{s}_{\mathcal{A}_{\mathrm{md}}}|\mathbf{s}_{\mathcal{A}_{\mathrm{eq}}},\mathbf{H})\big]^{2}}
  {4c^{2}+2c\delta_{2}I(\mathbf{y};\mathbf{s}_{\mathcal{A}_{\mathrm{md}}}|\mathbf{s}_{\mathcal{A}_{\mathrm{eq}}},\mathbf{H})}
  -\log k_{\ell}\\
  &\qquad \quad-i\log(k_{\ell}/i)-(k_{\ell}-i)\log\big[k_{\ell}/(k_{\ell}-i)\big].
\end{align}
The mutual information is calculated as
\begin{align}\label{B2}
  I(\mathbf{y};\mathbf{s}_{\mathcal{A}_{\mathrm{md}}}|\mathbf{s}_{\mathcal{A}_{\mathrm{eq}}},\mathbf{H})
  &=\mathbb{E}_{\mathbf{H}}\Big\{\log\det\Big(\mathbf{I}_{N_{R}}+\sum_{k \in \mathcal{A}_{\mathrm{md}}}\mathbf{H}_{k}\mathbf{Q}_{k}\mathbf{H}^{\dagger}_{k}\Big)\Big\},
\end{align}
When $i=|\mathcal{A}_{\mathrm{md}}|$ is finite, so does the mutual information. It then follows that~\eqref{B1} is on the order of
\begin{equation}\label{B3}
\Omega\big(\kappa n_{0}-(i+1)\log k_{\ell}\big),
\end{equation}
where the constant $\kappa$ is given by
\begin{equation}\label{B4}
  \kappa = \frac{\delta_{2}I(\mathbf{y};\mathbf{s}_{\mathcal{A}_{\mathrm{md}}}|\mathbf{s}_{\mathcal{A}_{\mathrm{eq}}},\mathbf{H})}
  {4c^{2}/(\delta_{2}I(\mathbf{y};\mathbf{s}_{\mathcal{A}_{\mathrm{md}}}|\mathbf{s}_{\mathcal{A}_{\mathrm{eq}}},\mathbf{H}))
  +2c}.
\end{equation}
On the other hand, when $i=|\mathcal{A}_{\mathrm{md}}|$ is finite, the R.H.S. of~\eqref{48} is on the order of
\begin{equation}\label{B5}
  \Omega\big(i\log(\ell-k_{\ell})+(1+1/\epsilon')i\log k_{\ell}+\log k_{\ell}\big).
\end{equation}
Combining~\eqref{B5} and~\eqref{B3}, it can be observed that by choosing $\delta_{2}\rightarrow 0$ sufficient slowly and with a sufficiently large implied constant, the condition in~\eqref{48} renders~\eqref{B3} growing unbounded with $\ell$, hence establishes the condition~\eqref{B1}. Further, by the assumption $\ell\gg k_{\ell}$,~\eqref{B5} is dominated by the first term. The others can be factorized into the constant $\epsilon$, and hence we get the final result in~\eqref{11}.

\emph{Case b)}: $i=|\mathcal{A}_{\mathrm{md}}|\leq k_{\ell}$ but grows unbounded with $\ell$. In this case, with the mutual information $I(\mathbf{y};\mathbf{s}_{\mathcal{A}_{\mathrm{md}}}|\mathbf{s}_{\mathcal{A}_{\mathrm{eq}}},\mathbf{H})>0$, the terms in~\eqref{B1} behaves as
\begin{equation}\label{B6}
  \Theta\Big(\kappa n_{0}I(\mathbf{y};\mathbf{s}_{\mathcal{A}_{\mathrm{md}}}
  |\mathbf{s}_{\mathcal{A}_{\mathrm{eq}}},\mathbf{H})
  -\log k_{\ell}-\log\binom{k_{\ell}}{i}\Big),
\end{equation}
for some constant $\kappa=\delta_{2}/\big[4c^{2}/(\delta_{2}I(\mathbf{y};\mathbf{s}_{\mathcal{A}_{\mathrm{md}}}|\mathbf{s}_{\mathcal{A}_{\mathrm{eq}}},\mathbf{H}))
  +2c\big]$.

On the other hand, by the assumption $\ell\gg k_{\ell} \geq i$, the numerator of $n_{0}$ in~\eqref{48} is dominated by the first term. Thus, we can factorize the terms $\frac{1}{\epsilon'}\log\binom{k_{\ell}}{i}-\log\delta_{1}+\frac{\gamma}{\epsilon'}$ into the constant $\epsilon$. With $i=|\mathcal{A}_{\mathrm{md}}|$ growing unbounded with $\ell$, the R.H.S. of $n_{0}$ in~\eqref{48} is now on the order of
\begin{equation}\label{B7}
  \Theta\Bigg(\frac{\log\binom{\ell-k_{\ell}}{i}+\log\binom{k_{\ell}}{i}+\log k_{\ell}}
  {I(\mathbf{y};\mathbf{s}_{\mathcal{A}_{\mathrm{md}}}|\mathbf{s}_{\mathcal{A}_{\mathrm{eq}}},\mathbf{H})}\Bigg).
\end{equation}
Combining~\eqref{B7} and~\eqref{B6}, it can be concluded that by choosing $\delta_{2}\rightarrow 0$ sufficient slowly and with a sufficiently large implied constant, the condition in~\eqref{48} renders~\eqref{B6} growing unbounded with $\ell$, hence establishes the condition~\eqref{47}. Further, by factorizing the last two terms in~\eqref{B7} into the constant $\epsilon$, we obtain the final result in~\eqref{11}.

\section{Proof of (65)}
Without loss of generality, we assume $\mathcal{A}_{l}=\{1,2,\ldots,\tilde{k}\}$. Due to the i.i.d. generation of the codewords and the memoryless nature channels, we have
\begin{align}\label{C1}
   E_{o}(\rho,s,\mathcal{A}_{l})=-n\log\mathbb{E}_{\mathbf{H}}\Bigg\{\int Q(\check{\mathbf{s}}_{\mathcal{A}_{l}^{c}})
   \Bigg[\int  Q(\check{\mathbf{s}}_{\mathcal{A}_{l}})
  P(\check{\mathbf{y}}|\check{\mathbf{s}}_{\mathcal{A}},\mathbf{H})^{\frac{1}{1+\rho}}d\check{\mathbf{s}}_{\mathcal{A}_{l}} \Bigg]^{1+\rho} d\check{\mathbf{s}}_{\mathcal{A}_{l}^{c}}d\check{\mathbf{y}}\Bigg\},
\end{align}
where $\check{\mathbf{s}}_{k} \in \mathbb{C}^{N_{T}\times 1}$ denote transmit codewords in each channel use, and the received signal is written as
\begin{equation}\label{C2}
  \check{\mathbf{y}}=\sum_{k\in \mathcal{A}_{l}}\mathbf{H}_{k}\check{\mathbf{s}}_{k}+\mathbf{z},
\end{equation}
with the channel matrix $\mathbf{H}_{k}\in \mathbb{C}^{N_{R}\times N_{T}}$.

Since the codeword $\mathbf{s}_{k}$ follows $\mathcal{CN}(\mathbf{0},\mathbf{Q}_{k})$, the integration within~\eqref{C1} can be calculated as
\begin{align}\label{C4}
  \nonumber&\int  Q(\check{\mathbf{s}}_{\mathcal{A}_{l}})
 P(\check{\mathbf{y}}|\check{\mathbf{s}}_{\mathcal{A}},\mathbf{H})^{\frac{1}{1+\rho}}d\check{\mathbf{s}}_{\mathcal{A}_{l}}\\
 \nonumber=&\int_{\check{\mathbf{s}}_{\mathcal{A}_{l}}} \frac{1}{\pi^{N_{R}\tilde{k}}\prod_{k\in \mathcal{A}_{l}}\det(\mathbf{Q}_{k})}\exp\Big\{-\sum_{k \in \mathcal{A}_{l}}\check{\mathbf{s}}_{k}^{\dagger}\mathbf{Q}_{k}^{-1}\check{\mathbf{s}}_{k}\Big\}\\
&\qquad\times\frac{1}{\pi^{N_{R}/(1+\rho)}}\exp\Big\{-\frac{1}{1+\rho}\Big\|\check{\mathbf{y}}-\sum_{k\in \mathcal{A}}\mathbf{H}_{k}\check{\mathbf{s}}_{k}\Big\|^{2}\Big\}.
\end{align}
Let $\check{\mathbf{y}}_{\mathcal{A}_{l}^{c}}=\check{\mathbf{y}}-\sum_{k\in \mathcal{A}_{l}^{c}}\mathbf{H}_{k}\check{\mathbf{s}}_{k}$,
\begin{equation}\label{C5}
  \mathbf{s}_{\mathcal{A}_{l}}= \begin{bmatrix}
                     \mathbf{s}_{1} \\
                     \mathbf{s}_{2} \\
                     \cdots \\
                     \mathbf{s}_{\tilde{k}} \\
                   \end{bmatrix}_{\tilde{k}N_{T}\times 1},
 \mathbf{Q}_{\mathcal{A}_{l}}=\begin{bmatrix}
   \mathbf{Q}_{1} &    &  \\
    &  \ddots &  \\
    &  &    \mathbf{Q}_{\tilde{k}} \\
 \end{bmatrix}_{\tilde{k}N_{T}\times \tilde{k}N_{T}},
\end{equation}
and
\begin{equation}\label{C6}
  \mathbf{H}_{\mathcal{A}_{l}}=\big[\mathbf{H}_{1},\mathbf{H}_{2},\cdots,\mathbf{H}_{\tilde{k}}\big]_{N_{R}\times\tilde{k}N_{T}}.
\end{equation}
Eq.~\eqref{C4} can be written as
\begin{align}\label{C7}
  \nonumber&\int  Q(\check{\mathbf{s}}_{\mathcal{A}_{l}})
  P(\check{\mathbf{y}}|\check{\mathbf{s}}_{\mathcal{A}},\mathbf{H})^{\frac{1}{1+\rho}}d\check{\mathbf{s}}_{\mathcal{A}_{l}}\\
  =&\int_{\check{\mathbf{s}}_{\mathcal{A}_{l}}}\frac{1}{\pi^{N_{T}\tilde{k}}\det(\mathbf{Q}_{\mathcal{A}_{l}})}
 \exp\{-\mathbf{s}_{\mathcal{A}_{l}}^{\dag}\mathbf{Q}_{\mathcal{A}_{l}}^{-1}\mathbf{s}_{\mathcal{A}_{l}}\}\frac{1}{\pi^{N_{R}/(1+\rho)}}\exp\Big\{-\frac{1}{1+\rho}\big\|\check{\mathbf{y}}_{\mathcal{A}_{l}^{c}}-\mathbf{H}_{\mathcal{A}_{l}}\mathbf{s}_{\mathcal{A}_{l}}\big\|^{2}\Big\}\\
 \nonumber=&\,\frac{1}{\pi^{N_{T}\tilde{k}}\det(\mathbf{Q}_{\mathcal{A}_{l}})}\frac{1}{\pi^{N_{R}/(1+\rho)}}\int_{\check{\mathbf{s}}_{\mathcal{A}_{l}}}
 \exp\Big\{-\mathbf{s}_{\mathcal{A}_{l}}^{\dag}\mathbf{G}^{-1}_{\mathcal{A}_{l}}\mathbf{s}_{\mathcal{A}_{l}}\\
 &\qquad \qquad \qquad -\frac{1}{1+\rho}\Big[\check{\mathbf{y}}_{\mathcal{A}_{l}^{c}}^{\dag}\mathbf{H}_{\mathcal{A}_{l}}\mathbf{s}_{\mathcal{A}_{l}}
+(\mathbf{H}_{\mathcal{A}_{l}}\mathbf{s}_{\mathcal{A}_{l}})^{\dag}\check{\mathbf{y}}_{\mathcal{A}_{l}^{c}}+\|\check{\mathbf{y}}_{\mathcal{A}_{l}^{c}}\|^{2}\Big]\Big\}\\
\nonumber=&\,\frac{\det(\mathbf{G}_{\mathcal{A}_{l}})}{\det(\mathbf{Q}_{\mathcal{A}_{l}})}\frac{1}{\pi^{N_{R}/(1+\rho)}}\int_{\check{\mathbf{s}}_{\mathcal{A}_{l}}}\frac{1}{\pi^{N_{T}\tilde{k}}\det(\mathbf{G}_{\mathcal{A}_{l}})}\\
&\qquad\times\exp\Big\{-\big(\mathbf{s}_{\mathcal{A}_{l}}+\pmb{\mu}_{\mathcal{A}_{l}}\big)^{\dag}\mathbf{G}_{\mathcal{A}_{l}}^{-1}\big(\mathbf{s}_{\mathcal{A}_{l}}+\pmb{\mu}_{\mathcal{A}_{l}}\big)\Big\}\exp\Big\{-\frac{1}{1+\rho}\|\check{\mathbf{y}}_{\mathcal{A}_{l}^{c}}\|^{2}+\pmb{\mu}_{\mathcal{A}_{l}}^{\dag}\mathbf{G}_{\mathcal{A}_{l}}^{-1}\pmb{\mu}_{\mathcal{A}_{l}}\Big\}\\
=&\,\frac{\det(\mathbf{G}_{\mathcal{A}_{l}})}{\det(\mathbf{Q}_{\mathcal{A}_{l}})}\frac{1}{\pi^{N_{R}/(1+\rho)}}\exp\Big\{-\frac{1}{1+\rho}\check{\mathbf{y}}_{\mathcal{A}_{l}^{c}}^{\dag}\Big[\mathbf{I}_{N_{R}}
-\frac{1}{1+\rho}\mathbf{H}_{\mathcal{A}_{l}}\mathbf{G}_{\mathcal{A}_{l}}\mathbf{H}_{\mathcal{A}_{l}}^{\dag}\Big]\check{\mathbf{y}}_{\mathcal{A}_{l}^{c}}\Big\},
\end{align}
where
\begin{equation}\label{C8}
  \mathbf{G}_{\mathcal{A}_{l}}=\Big(\mathbf{Q}_{\mathcal{A}_{l}}^{-1}+\frac{1}{1+\rho}\mathbf{H}^{\dag}_{\mathcal{A}_{l}}\mathbf{H}_{\mathcal{A}_{l}}\Big)^{-1},
\end{equation}
and
\begin{equation}\label{C9}
  \pmb{\mu}_{\mathcal{A}_{l}}=\frac{1}{1+\rho}\mathbf{G}_{\mathcal{A}_{l}}\mathbf{H}_{\mathcal{A}_{l}}^{\dag}\check{\mathbf{y}}_{\mathcal{A}_{l}^{c}}.
\end{equation}

Let
\begin{align}\label{C10}
  \mathbf{\Omega}_{\mathcal{A}_{l}}&=\Big[\mathbf{I}_{N_{R}}-\frac{1}{1+\rho}\mathbf{H}_{\mathcal{A}_{l}}\mathbf{G}_{\mathcal{A}_{l}}\mathbf{H}_{\mathcal{A}_{l}}^{\dag}\Big]^{-1}\\
&=\mathbf{I}_{N_{R}}+\frac{1}{1+\rho}\mathbf{H}_{\mathcal{A}_{l}}\mathbf{Q}_{\mathcal{A}_{l}}\mathbf{H}_{\mathcal{A}_{l}}^{\dag}\\
&=\mathbf{I}_{N_{R}}+\frac{1}{1+\rho}\sum_{k \in \mathcal{A}_{l}}\mathbf{H}_{k}\mathbf{Q}_{k}\mathbf{H}^{\dagger}_{k},
\end{align}
it then follows that~\eqref{C1} can be calculated as
\begin{align}\label{C11}
  E_{o}(\rho,\mathcal{A}_{l})&=-n\log\mathbb{E}_{\mathbf{H}}\Bigg\{\det(\mathbf{\Omega}_{\mathcal{A}_{l}})
\Bigg[\frac{\det(\mathbf{G}_{\mathcal{A}_{l}})}{\det(\mathbf{Q}_{\mathcal{A}_{l}})}\Bigg]^{1+\rho}\Bigg\}\\
&=-n\log\mathbb{E}_{\mathbf{H}}\Big\{
  \det\Big(\mathbf{I}_{N_{R}}+\frac{1}{1+\rho}\mathbf{G}_{\mathcal{A}_{l}}\Big)^{-\rho}\Big\},
\end{align}
where $\mathbf{G}_{\mathcal{A}_{l}}=\sum \limits_{k \in \mathcal{A}_{l}}
  \mathbf{H}_{k}\mathbf{Q}_{k}\mathbf{H}^{\dagger}_{k}$.

\section{Proof of Proposition 2}
We choose $\rho=1$ in the error exponent $E_{r}(\rho, \mathcal{A}_{l})$. By combining~\eqref{60},~\eqref{61} and noting that $\log\binom{k_{n}}{\tilde{k}}\leq k_{n}H_{2}(\tilde{k}/k_{n}) $,  $E_{r}(\rho, \mathcal{A}_{l})$ can be lower bounded as
\begin{align}\label{D1}
\nonumber E_{r}(\rho, \mathcal{A}_{l})|_{\rho=1} &\geq -\frac{k_{n}}{n}H_{2}(\tilde{k}/k_{n}) -\log\mathbb{E}_{\mathbf{H}}\Big\{\det\Big(\mathbf{I}_{N_{R}}+\frac{1}{2}\sum_{t \in \mathcal{A}_{l}}\mathbf{H}_{t}\mathbf{Q}_{t}\mathbf{H}^{\dagger}_{t}\Big)^{-1}\Big\}\\
  &-\frac{(1-\epsilon)}{n}\sum_{k \in \mathcal{A}_{l}}
  c_{k}\mathbb{E}_{\mathbf{H}}\Big\{\log\det\big(\mathbf{I}_{N_{R}}+\mathbf{G}_{\mathcal{A}}\big)\Big\},
\end{align}
where $\mathbf{G}_{\mathcal{A}}=\sum_{t \in \mathcal{A}}\mathbf{H}_{t}\mathbf{Q}_{t}\mathbf{H}^{\dagger}_{t}$.
In the following, we consider the separated cases and show that $E_{r}(\rho, \mathcal{A}_{l}) > 0$ as $n\rightarrow\infty$.

\emph{Case a)}: $\tilde{k}$ scales linearly with codelength $n$ such that $\lim\limits_{n\rightarrow\infty}\tilde{k}/k_{n}=\gamma$, where $\gamma>0$ is some constant. Since $\sum_{k \in \mathcal{A}_{l}}c_{k} < n$ and $H_{2}(\tilde{k}/k_{n})<1$, we have
\begin{align}\label{D2}
\nonumber E_{r}(\rho, \mathcal{A}_{l})|_{\rho=1} &\geq \epsilon\mathbb{E}_{\mathbf{H}}\Big\{\log\det\big(\mathbf{I}_{N_{R}}+\sum_{t \in \mathcal{A}}\mathbf{H}_{t}\mathbf{Q}_{t}\mathbf{H}^{\dagger}_{t}\big)\Big\}\\ \nonumber -\frac{k_{n}}{n}
-\Bigg\{\log&\,\mathbb{E}_{\mathbf{H}}\Big\{\det\Big(\mathbf{I}_{N_{R}}+\frac{1}{2}\sum_{t \in \mathcal{A}_{l}}\mathbf{H}_{t}\mathbf{Q}_{t}\mathbf{H}^{\dagger}_{t}\Big)^{-1}\Big\}\\
+\mathbb{E}_{\mathbf{H}}&\Big\{\log\det\big(\mathbf{I}_{N_{R}}+\sum_{t \in \mathcal{A}}\mathbf{H}_{t}\mathbf{Q}_{t}\mathbf{H}^{\dagger}_{t}\big)\Big\}\Bigg\}.
\end{align}
Define matrix $\mathbf{G}_{t} = \mathbf{H}_{t}\mathbf{Q}_{t}\mathbf{H}^{\dagger}_{t}$, where the $(i,j)_{th}$ entry is given by
\begin{align}\label{D3}
   \nonumber g_{i,j}^{(t)} &= \sum_{n}\sum_{m}h_{i,m}^{(t)}q_{m,n}^{(t)}(h_{j,n}^{(t)})^{*} \\
   &=\sum_{n}\sum_{m}\alpha_{i,m}^{(t)}q_{m,n}^{(t)}(\alpha_{j,n}^{(t)})^{*}\beta_{t}.
\end{align}
Since $\alpha_{i,m}^{(t)}$ follows standard normal distribution and $|\mathcal{A}|=\tilde{k}=O(n)$, by Kolmogorov strong law of large number~\cite{Sen93}, we have
\begin{equation}\label{D4}
   \lim \limits_{k_{n} \rightarrow \infty} \sum\limits_{t \in \mathcal{A}}g_{i,j}^{(t)} = \left\{
      \begin{array}{ll}
       \sum\limits_{t \in \mathcal{A}}\beta_{t}\mathrm{Tr}(\mathbf{Q}_{t}), & \hbox{$i = j$;} \\
         0, & \hbox{$i \neq j$.}
       \end{array}
   \right.
\end{equation}
and
\begin{equation}\label{D5}
  \lim \limits_{\tilde{k} \rightarrow \infty} \sum\limits_{t \in \mathcal{A}_{l}}g_{i,j}^{(t)} = \left\{
        \begin{array}{ll}
          \sum\limits_{t \in \mathcal{A}_{l}}\beta_{t}\mathrm{Tr}(\mathbf{Q}_{t}), & \hbox{$i = j$;} \\
           0, & \hbox{$i \neq j$.}
        \end{array}
        \right.
\end{equation}

Consequently,
\begin{equation}\label{D6}
      \lim \limits_{k_{n}\rightarrow \infty} \det\Big(\mathbf{I}_{N_{R}}+\sum \limits_{t \in \mathcal{A}}\mathbf{G}_{t}\Big)
      = \Big(1+  \sum\limits_{t \in \mathcal{A}}\beta_{t}\mathrm{Tr}(\mathbf{Q}_{t})\Big)^{N_{R}},
\end{equation}
and
\begin{equation}\label{D7}
    \lim \limits_{\tilde{k} \rightarrow \infty} \det\Big(\mathbf{I}_{N_{R}}+\frac{1}{2}\sum \limits_{t \in \mathcal{A}_{l}}\mathbf{G}_{t}\Big)
    =\Big(1+\frac{1}{2}\sum\limits_{t \in \mathcal{A}_{l}}\beta_{t}\mathrm{Tr}(\mathbf{Q}_{t})\Big)^{N_{R}}.
\end{equation}
Given~\eqref{D6} and~\eqref{D7}, the last two term in~\eqref{D2} converge to some constant with finite number of receiving antennas $N_{R}$ and constant power $p_{k}=\Theta(1)$, i.e., $\lim_ {n\rightarrow\infty} k_{n}/n=\Theta(1)$ and
\begin{align}\label{D8}
   \nonumber &\lim_{n\rightarrow\infty}\log \mathbb{E}_{\mathbf{H}} \Big\{
   \det\Big(\mathbf{I}_{N_{R}}+\frac{1}{2}\sum_{t \in
   \mathcal{A}_{l}}\mathbf{H}_{t}\mathbf{Q}_{t}\mathbf{H}^{\dagger}_{t}\Big)^{-1}\Big\}\\
   \nonumber &\qquad \qquad \qquad+\mathbb{E}_{\mathbf{H}}\Big\{\log\det\Big(\mathbf{I}_{N_{R}}+\sum_{t \in \mathcal{A}}\mathbf{H}_{t}\mathbf{Q}_{t}\mathbf{H}^{\dagger}_{t}\Big)\Big\}\\
   &=N_{R}\log\Bigg[\Big(1+\sum\limits_{t \in \mathcal{A}}\beta_{t}p_{t}\Big)\Big/\Big(1+\frac{1}{2}\sum\limits_{t \in \mathcal{A}_{l}}\beta_{t}p_{t}\Big)\Bigg]\\
   &=N_{R}O\big(\log(2/\gamma)\big).
\end{align}

On the other hand, the first term in~\eqref{D2} grows at a speed $O(\log(k_{n}))$, i.e.,
\begin{align}\label{D9}
       \nonumber &\lim_{n\rightarrow\infty} \epsilon \mathbb{E}_{\mathbf{H}}\Big\{\log\det\Big(\mathbf{I}_{N_{R}}+\sum_{t \in \mathcal{A}}\mathbf{H}_{t}\mathbf{Q}_{t}\mathbf{H}^{\dagger}_{t}\Big)\Big\}\\
       &=\epsilon N_{R}\log\Big(1+\sum\limits_{t \in \mathcal{A}}\beta_{t}p_{t}\Big)\\
       &=\epsilon N_{R}O(\log k_{n}).
\end{align}
Thus, we have $E_{r}(1,\mathcal{A}_{l})>c_{0}$ as $n\rightarrow\infty$.

\emph{Case b)}: $\tilde{k}$ scales sublinearly with codelength $n$ such that $\lim_{n\rightarrow\infty}\tilde{k}/k_{n}=0$. Since $\lim\limits_{n\rightarrow\infty}H_{2}(\tilde{k}/k_{n})=0$, the first term in~\eqref{D1} vanishes as $n\rightarrow\infty$. Note that $\mathbf{H}_{t}\mathbf{Q}_{t}\mathbf{H}^{\dagger}_{t}$ is positive semi-definite, its largest eigenvalue is strictly positive, and hence the largest eigenvalue of $\mathbf{I}_{N_{R}}+\frac{1}{2}\sum_{t \in \mathcal{A}_{l}}\mathbf{H}_{t}\mathbf{Q}_{t}\mathbf{H}^{\dagger}_{t}$ is strictly larger than $1$, so that the second term in~\eqref{D1} is strictly positive. As $\tilde{k}$ grows sublinearly with $n$, the last two terms behaves as
\begin{equation}\label{D10}
   - \log \mathbb{E}_{\mathbf{H}}\Big\{\det\Big(\mathbf{I}_{N_{R}}+\frac{1}{2}
   \sum_{t \in \mathcal{A}_{l}}\mathbf{H}_{t}\mathbf{Q}_{t}\mathbf{H}^{\dagger}_{t}\Big)^{-1}\Big\} = O(\log\tilde{k}),
\end{equation}
and
\begin{align}\label{D11}
    \frac{(1-\epsilon)}{n}\sum_{k \in\mathcal{A}_{l}}c_{k}
    \mathbb{E}_{\mathbf{H}}\Big\{\log\det\big(\mathbf{I}_{N_{R}}+\sum_{t \in \mathcal{A}}
    \mathbf{H}_{t}\mathbf{Q}_{t}\mathbf{H}^{\dagger}_{t}\big)\Big\}=O\Big(\frac{\tilde{k}}{n}\log k_{n}\Big),
\end{align}
respectively.

Since $\lim\limits_{n\rightarrow\infty}\tilde{k}/k_{n}=0$ and $k_{n}=O(n)$, we have the following asymptotic inequality
\begin{equation}\label{D12}
   \log\tilde{k} > \frac{\tilde{k}}{n}\log k_{n},
\end{equation}
as $n\rightarrow\infty$. Hence establish the proof.
\section*{Acknowledgment}

The authors would like to thank...

\ifCLASSOPTIONcaptionsoff
  \newpage
\fi

\end{document}